\documentclass[pdflatex,sn-mathphys-num]{sn-jnl}
\usepackage{graphicx}%
\usepackage{multirow}%
\usepackage{amsmath,amssymb,amsfonts}%
\usepackage{amsthm}%
\usepackage{mathrsfs}%
\usepackage[title]{appendix}%
\usepackage{xcolor}%
\usepackage{textcomp}%
\usepackage{manyfoot}%
\usepackage{booktabs}%
\usepackage{algorithm}%
\usepackage{algorithmicx}%
\usepackage{algpseudocode}%
\usepackage{listings}%
\usepackage{subcaption}
\usepackage{placeins}
\usepackage[font=small]{caption}

\theoremstyle{thmstyleone}%
\theoremstyle{thmstyletwo}%

\theoremstyle{thmstylethree}%

\begin{document}

\title[Strategy-first synthesis planning for complex natural products]{Strategy-first synthesis planning for complex natural products}

\author[1]{\fnm{Daniel} \sur{Armstrong}}\email{daniel.armstrong@epfl.ch}
\equalcont{These authors contributed equally to this work.}
\author[1]{\fnm{Xuan-Vu} \sur{Nguyen}}\email{nguyen.nguyen@epfl.ch}
\equalcont{These authors contributed equally to this work.}
\author[1]{\fnm{Octavian} \sur{Susanu}}
\author[1]{\fnm{Gabriel} \sur{Gibberd}}
\author[1]{\fnm{Théo A.} \sur{Neukomm}}
\author[1]{\fnm{Taddäus} \sur{Strunden}}
\author[2]{\fnm{Dan} \sur{Forster}}
\author[2]{\fnm{Morgane} \sur{Delattre}}
\author[2]{\fnm{Shawn} \sur{Teh}}
\author[2,3]{\fnm{Clément} \sur{Rols}}
\author[5]{\fnm{John} \sur{Federice}}
\author[5]{\fnm{Hayden} \sur{Leatherwood}}
\author[6]{\fnm{M. Lavelle} \sur{Barnes}}
\author[1,4]{\fnm{Maarten R.} \sur{Dobbelaere}}
\author[6]{\fnm{Peter} \sur{Wipf}}\email{pwipf@pitt.edu}
\author[5]{\fnm{Jon T.} \sur{Njardarson}}\email{njardars@arizona.edu}
\author[2,3]{\fnm{Jieping} \sur{Zhu}}\email{jieping.zhu@epfl.ch}
\author*[1,3]{\fnm{Philippe} \sur{Schwaller}}\email{philippe.schwaller@epfl.ch}

\affil[1]{\orgdiv{Laboratory of Artificial Chemical Intelligence (LIAC)}, \orgname{\'Ecole Polytechnique F\'{e}d\'{e}rale de Lausanne (EPFL)}, \orgaddress{\city{Lausanne}, \country{Switzerland}}}

\affil[2]{\orgdiv{Laboratory of Synthesis and Natural Products (LSPN)}, \orgname{\'Ecole Polytechnique F\'{e}d\'{e}rale de Lausanne (EPFL)}, \orgaddress{\city{Lausanne}, \country{Switzerland}}}

\affil[3]{\orgname{National Centre of Competence in Research (NCCR) Catalysis}, \orgaddress{\city{Lausanne}, \country{Switzerland}}}

\affil[4]{\orgname{Laboratory for Chemical Technology, Ghent University}, \orgaddress{\city{Zwijnaarde}, \country{Belgium}}}

\affil[5]{\orgdiv{Njardarson Laboratory}, \orgname{University of Arizona}, \orgaddress{\city{Tucson}, \country{United States}}}

\affil[6]{\orgdiv{Wipf Group}, \orgname{University of Pittsburgh}, \orgaddress{\city{Pittsburgh}, \country{United States}}}

\abstract{The total synthesis of a complex molecule is among the most demanding intellectual and experimental feats in chemistry: a chemist must plan many steps ahead for how to assemble simple building blocks into an intricate target, devise backup strategies, and anticipate procedural challenges. It is also a profoundly creative activity. For half a century, efforts to automate the retrosynthetic design of natural products and other complex molecules have drawn on catalogued reactions, and the resulting tools now report near-complete success on benchmarks built from that same source. But these tools were shaped to fit benchmarked chemistry, and they falter on many natural products, the frontier of the field, whose densely functionalized, polycyclic architectures demand precisely the inventive chemistry the record contains least. Whether a machine could reasonably design such syntheses like an expert chemist does has remained unclear. Here, we show that SynthEx, an agentic framework built on large language models, plans routes to complex natural products that lie beyond the reach of conventional design algorithms. Freed from the fixed reaction libraries that confine those planners, SynthEx proposes competing strategies, assembles a sequence of routine and key steps into a cohesive route, and critiques and improves its own design; the chemistry it favors is more convergent and elegant than existing tools produce, and spans a region of reaction space that catalogue-based tools cannot match. Most notably, in blinded assessments, expert chemists judged its key steps comparable to those of published human syntheses and engaged with them as genuine synthesis plans, a response algorithmic route prediction has not previously accomplished. We release routes to more than a thousand natural products as SynthAtlas, an open, interactive roadmap, and anticipate it will become a shared resource for a collection of complex target molecules that lack relevant current literature designs.}




\maketitle

\section{Main}\label{sec1}

The total synthesis of natural products is among the most challenging and creative endeavors in chemistry. Faced with an intricate molecular architecture and complex stereochemical relationships, a chemist develops a reaction-by-reaction recipe (synthetic route), deciding which bonds to forge and in what order, when to introduce and remove protective groups, and how to control the three-dimensional outcome of every step to create the target molecule. Corey's retrosynthetic analysis gave this reasoning a formal language, reducing a formidable target to stepwise disconnections, resulting in a sequence of established, feasible transformations \citep{corey1964total, Corey_Cheng_LogicofChemicalSynthesis, nicolaou1996classics, nicolaou2000}.  The discipline has repaid the effort many times over: total synthesis is where new reactions prove their worth and many, now-standard, transformations trace their origins to the assembly of a natural product \cite{trost1991comprehensive,nicolaou1997wittig,heravi2021recent}. Yet designing a route to a demanding target remains an art that only a small community of experts practice fluently \citep{bran2026chemical}, and it is that expertise, not the chemistry it draws on, that proves hardest to articulate and imitate.

For half a century, chemists have sought to automate synthesis planning. Computer-Assisted Synthesis Planning (CASP) began with Corey and Wipke in the 60s \citep{corey1969computer, corey1972computer, corey1985computer}, with manually coded reaction logic, before shifting to reaction templates—deterministic rules mined at scale from literature and patents \citep{lowe2011chemical, lowe2012extraction, Lowe2017}. Trained retrosynthesis models can now propose multistep routes in minutes, and on standard benchmarks drawn from these same corpora, they succeed impressively \citep{Segler2017neural, coley2017computer, segler2018planning, Schwaller2018found, Jin2017predicting,chen2020retro, Sacha2021molecule, schwaller2020predicting,pdvn, maziarz2024chimera}. However, this success reflects the shared distribution of the benchmarks and training data, which are dominated by the recurring, routine transformations of streamlined medicinal and process chemistry. The highly context-specific disconnections that complex targets demand are precisely what these tools struggle to capture \citep{3aizynthtrain, schwaller2019molecular, tanovic2026exploration}.

This limitation has become visible in two ways. First, 
the benchmarks themselves have saturated: on the patent-derived test sets that the field has long used, state-of-the-art planners now report success approaching completeness. Like early coding benchmarks that saturated before the field shifted to agentic software engineering tasks \citep{jimenez2024swe,yang2024swe}, existing synthesis benchmarks no longer probe the axes on which model capabilities are improving \citep{genheden_bjerrum_2022,maziarz2023re,tripp2024retro,maziarz2024chimera,morgunov2025procrustean}. Second, and more tellingly, synthetic planning tools falter the moment they leave that chemical space \citep{xuan2025tempre, reactionspace_hassen,tran2026quantifying}. Complex natural products, with their densely fused, stereochemically rich architectures, are out-of-distribution for these tools. This weakness is likely structural to the reaction rule approach itself. Rules, whether as explicit templates, or implicit in neural network weights, rely on applying memorized patterns from frequently occurring reactions. Natural product synthesis, however, routinely requires bespoke, inventive transformations—such as complex ring-closing cascades or intricate structural rearrangements—that depend entirely on the specific, global context of the molecule. Because these reactions occur too rarely to form robust templates \citep{3aizynthtrain}, or are too rare to be recalled by a neural network \citep{schwaller2019molecular, maziarz2024chimera}, a planner confined to a historical reaction library fundamentally struggles to navigate the unique chemical environments of natural products. The field has accordingly moved past judging a planner on whether it can successfully retrieve a pathway, to whether a chemist would actually use the resulting route \citep{morgunov2026syntax}.

Large language models (LLMs) suggest a way past this limitation. Having absorbed, through pretraining, a vast body of chemical knowledge and context that extends well beyond any single reaction database, they can act less as retrieval engines than as chemical reasoning engines \citep{jablonka202314,jablonka2023gpt,alampara2024probing, bran2023chemcrow,boiko2023autonomous,mirza2025framework, armstrong2025synthstrategy}. Bran et al. showed that an LLM can judge a synthesis, aligning a chemist's stated strategy with candidate routes and ranking them in agreement with human experts \citep{bran2026chemical}. Frameworks such as LARC \cite{baker2025larc} and MMORF \cite{baker2026mmorf} use LLMs as evaluators, selecting and scoring nodes against user-defined criteria while a traditional planner expands the search space. Synthelite \cite{xuan2025synthelite} demonstrates that an LLM can directly plan, casting the model as a language-based policy prior that proposes each disconnection in natural language and so steers the tree search directly, in place of a trained reaction policy network; these proposals, however, are still grounded by retrieval against a fixed template library. Each advances the role an LLM can play, yet all share one ceiling: whether the model ranks synthetic strategy or tactics or proposes the next step, it selects within a reaction space a conventional planner has already fixed, and has no ability to instantiate reactions outside that space. This is what makes complex-molecule synthesis the most relevant available test of chemical reasoning: because the required disconnections are, by construction, decoupled from any reaction database, success is not limited to reaction retrieval, and a system that plans these routes must reason about reactivity, strategy, and stereocontrol rather than recall them through a template. Planning a synthesis in this way requires the planner to adapt chemistry principles to any given reaction and is a necessary precondition for the automation of chemical reasoning.

\begin{figure*}[htbp]
  \centering
  \includegraphics[width=\textwidth]{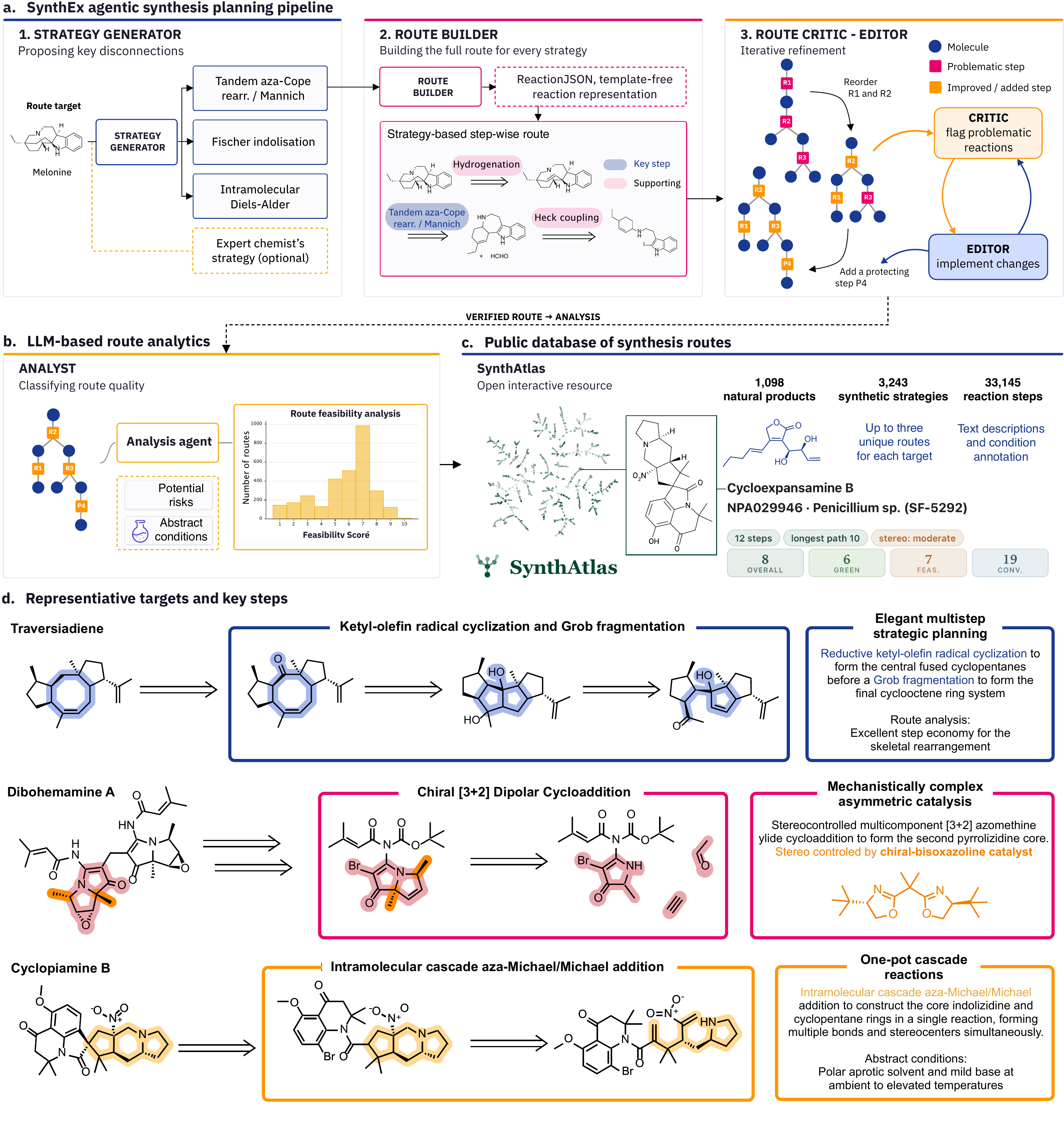}
  \caption{\textbf{The SynthEx agentic synthesis-planning pipeline and the
  SynthAtlas resource.}
  \textbf{a}, ~The planning pipeline has three stages with corresponding subagents. First, the \emph{Strategy Generator} proposes several
  high-level strategies for the target molecule, each centered on a key
  disconnection. In the second stage, the \emph{Route Builder}
  expands each strategy into a pathway, expressing disconnections
  in ReactionJSON, a template-free representation in which reactions are specified
  as graph-edit operations; key and supporting steps are distinguished.
  Next, the \textit{Critic} flags problematic steps
  (pink) and an \textit{Editor} applies edits to the route (orange), in an
  iterative loop that leaves the overall strategy intact.
  \textbf{b}, ~Verified routes are passed to an LLM-based analysis stage that
  classifies route quality, flagging potential risks and key reaction steps, and
  scoring feasibility; the histogram shows the distribution of route feasibility
  scores across the corpus.
  \textbf{c}, ~The routes are released as SynthAtlas, an open, interactive resource
  built over NP-Atlas targets. \textbf{d}, ~Representative key steps produced by SynthEx for the natural products Traversadiene, with a two-step cyclization followed by Grob fragmentation, Dibohemamine A, an asymmetric dipolar cycloaddition, and Cyclopiamine B, a cascade aza-Michael/Michael addition to close two rings in a single step.}
  \label{fig:pipeline}
\end{figure*}

 Here we introduce SynthEx, an agentic synthesis planner that removes this constraint at its source (Fig. \ref{fig:pipeline}). Rather than selecting disconnections from a fixed template library, SynthEx expresses each as an ordered list of atom-level graph edits, a format we call ReactionJSON. By letting the language model write these edits directly, SynthEx transitions from a critic that ranks known templates to a policy that generates novel expansions. Around this representation, SynthEx builds the iterative reasoning of a human chemist: it proposes competing high-level strategies, expands each into a complete pathway, and critiques and repairs its own chemistry step by step (Fig. \ref{fig:pipeline}a).

Applied to a collection of over a thousand documented natural products, SynthEx opens a region of synthesis space that a state-of-the-art search does not reach. It designs routes for the large majority of targets for which a leading template-based planner solves only a small fraction, and its advantage \emph{widens}, rather than narrows, as targets grow more complex: the additional rings, connectivities, and functional groups that cause a conventional search to fail have a lower impact on SynthEx. Its bond constructions are also markedly more convergent than those existing tools propose, most of them uniting two independent fragments, and they occupy a distinct reaction space that state-of-the-art single-step models rarely propose \citep{3aizynthtrain, maziarz2024chimera, lowe2011chemical, reaxys}. This is not the same chemistry produced in new contexts but different chemistry altogether: SynthEx favors constructive, bond-forming steps, ring constructions and cross-couplings, over the functional group manipulations that dominate template-based output, a qualitative signature of SynthEx's strategic preferences. None of the benchmark targets are flagged as synthesized in NPAtlas, so the routes SynthEx proposes are not retrievals of existing literature. In blinded review, expert chemists rated SynthEx's key steps on par with published human syntheses and, additionally, singled out individual disconnections, such as Grob fragmentation, [3+2] dipolar cycloaddition, or intramolecular cascade Michael additions, as genuinely elegant (Fig. \ref{fig:pipeline}d).
We release these routes as SynthAtlas (Fig. \ref{fig:pipeline}c), an open, interactive resource comprising 1,098 natural-product targets, 3,243 routes with corresponding synthetic strategies, and 33,145 reactions.

The focus of this work is on reach, convergence, distinctiveness, and strategy of proposed syntheses. We regard wet-lab feasibility as the next frontier in CASP, with the field moving from graph-based solve rate toward establishing proven chemical validity. To contribute to the automation of scientific discovery, every route is released in SynthAtlas, an open, interactive platform on which chemists can explore, compare, and comment on them. We envision that this atlas of predicted synthetic routes has the potential to spark ideas and discussions, and radically alter the way synthetic chemistry is approached, in analogy to how predicted protein structures transformed how biologists reason about molecules they may never crystallize \citep{jumper2021highly}.

\section{Results}\label{sec2}
\subsection{SynthEx: a multi-agent system for synthesis planning}\label{res:agent}
SynthEx plans a synthesis the way a chemist does: it settles on a strategy before it
commits to a route, and it revises the route without abandoning the strategy. Five
stages, each driven by a dedicated language-model agent, implement this
(Fig.~\ref{fig:pipeline}a). The \emph{Strategy Generator} agent proposes diverse
high-level strategies for the target, each anchored on a key disconnection. A
\emph{Route Builder} expands every strategy into a complete pathway, expressing each
disconnection in ReactionJSON, a template-free representation that we introduce here in
which a reaction is an ordered list of atom-level graph edits. A \emph{Critic} then simulates each
reaction in the forward direction and flags those that are chemically not feasible, and
an \emph{Editor} repairs them through surgical edits that leave the
strategy intact. Finally, an \emph{Analyst} scores the finished route for
feasibility and identifies its key steps and risks (Fig.~\ref{fig:pipeline}b). The
architecture descends from Synthelite \citep{xuan2025synthelite}, which treats a
language model as the search policy over a fixed template library while delegating
single-step feasibility to that same library. 

The first stage emphasizes strategic planning before reaction searching. Rather than commit to one disconnection, the
\textit{Strategy Generator} returns a diverse set of competing strategic hypotheses, three per
target by default, each built around a key step that the downstream stages then test
and prune. This mirrors the way a chemist entertains several plausible routes before
investing in one, and it frames synthesis planning as hypothesis generation followed
by exploration rather than as a single-shot search. Strategies can also be steered by
constraints expressed in natural language, a required starting material, or a
free-text instruction from a chemist, although for this work we leave the choice to
the model.

For the second stage, we design ReactionJSON, a reaction representation the model authors itself. Earlier LLM planners choose among entries in a template library, which serves complex reactions in unfamiliar structural contexts poorly; the obvious alternative, having the model generate reaction SMILES directly, is unreliable \citep{bran2026chemical,runcie2026molecular}. ReactionJSON avoids both bottlenecks: the model neither names a transformation nor redraws a molecule,
but writes the ordered graph edits that convert a product into its precursors;
applying those edits to the mapped product yields the precursors deterministically
(Methods~\ref{methods:synthex}). This converts the language model from a subroutine that selects from a limited number of available steps to a designer, and in doing so lifts the search off the reaction frequency and similarity precedence that limits every template-based and corpus-trained planner alike: SynthEx can propose chemistry that is common in the named-reaction and total-synthesis knowledge a model has absorbed but rare in any reaction catalogue, and therefore invisible to tools trained on one.

A practical artifact of ReactionJSON is that a route can be rendered as an editable object.
Because a ReactionJSON route is
a text object anchored on atom maps, it can be revised in place: the \textit{Critic} and \textit{Editor} can critique and repair the
chemistry surgically without re-running the search that produced it: reordering
reactions, inserting protections, or replacing a disconnection
(Section~\ref{res:improve}) without re-running the search that produced it 
(Section~\ref{res:improve}). A repair loop of this kind would be impractically expensive if 
each edit required a full tree search, and unreliable if SMILES needed regenerating by an LLM. 
This flexibility is what allows the system to genuinely correct its own chemistry rather than merely rank it.

For every target SynthEx returns several distinct strategies, each carrying step- and
route-level reasoning. Before quantifying reach across the full benchmark, we examine three demanding molecules in detail.

\subsection{Strategic synthesis planning and reasoning}

A solve rate cannot convey the essence of what a route actually proposes. Before turning to benchmark-scale reach (Section \ref{res:reach}), we examine the chemistry itself on three targets, ordered by how much external validation is available against which to check the proposals.. For Okaramine M a route was published after the model's training cut-off (Methods~\ref{methods:synthex}), and SynthEx recovers it.
Melonine presents a harder test: a total synthesis was likewise published after the cut-off, but here SynthEx does not converge on it, proposing instead an alternative disconnection that one of us (J.Z.), whose group has worked on this target, judges feasible and worthy of
experimental validation.
For the conversion of Chanoclavine to Lysergol, we are aware of no reported route across this specific gap: it is an open problem, posed here by an author working on it, on which the only available check is expert judgment of a novel proposal. In all three, what matters is not merely whether a route is found but whether the chemistry withstands a synthetic chemist's scrutiny.

\begin{figure*}[h!]
  \centering
  \includegraphics[width=\textwidth]{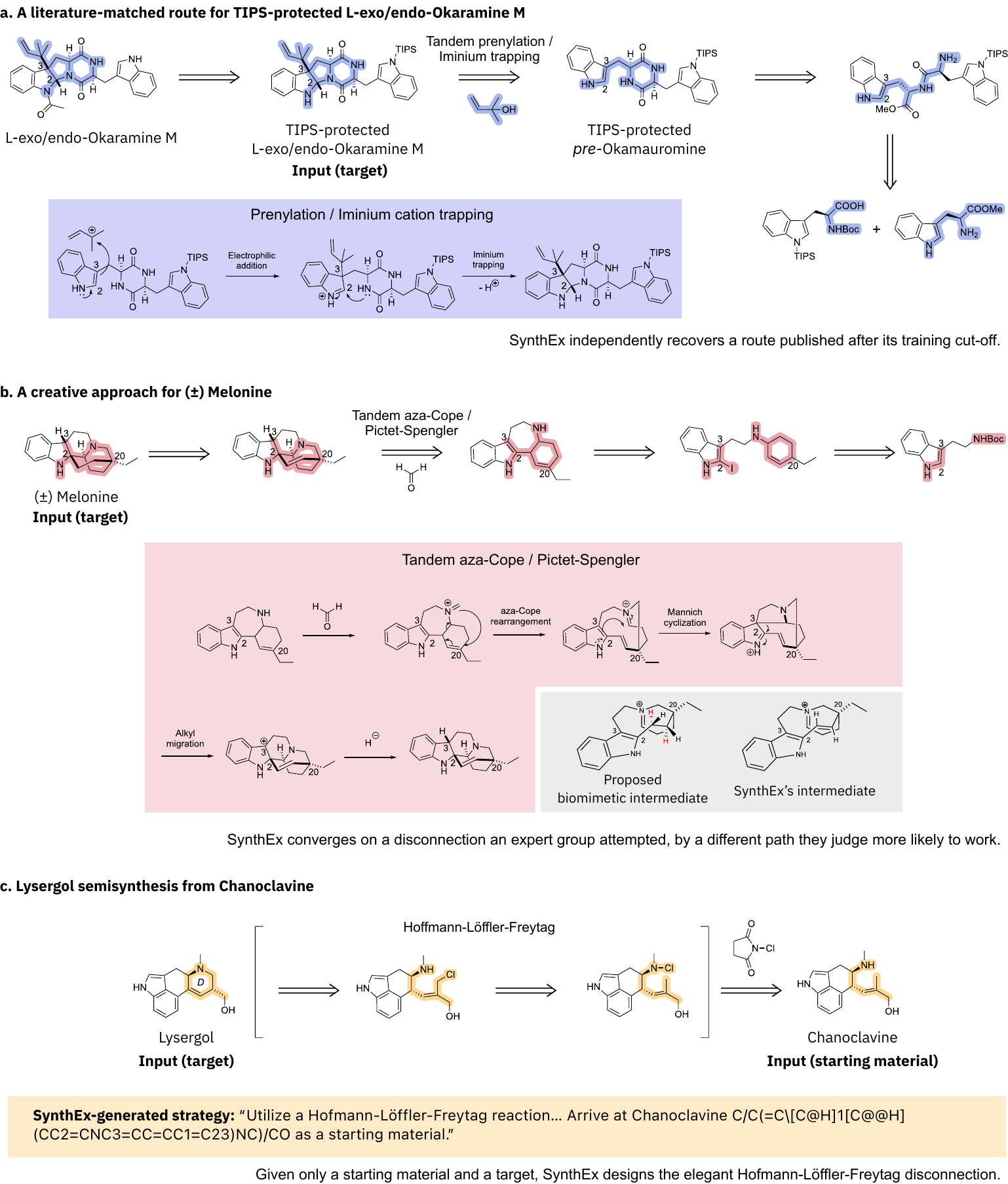}
  \caption{\textbf{Three case studies in strategic reasoning.} 
\textbf{a}, ~Okaramine~M, with the mechanism of the tandem prenylation and
  iminium trapping inset. \textbf{b}, ~Melonine, with the mechanism of the tandem
  aza-Cope / Pictet--Spengler key step.
  The gray inset compares the biomimetic pre-Mannich intermediate proposed in the literature with SynthEx's. The hydrogens causing potential steric clash with the piperidine ring are highlighted in red.
  \textbf{c}, ~Chanoclavine to Lysergol;
  the two structures shown were the entire input to the planner, which proposed the
  Hofmann--L\"{o}ffler--Freytag sequence unprompted.
  Colored atoms and bonds highlight the reaction center in the key step of each route.
  }
  \label{fig:casestudies}

\end{figure*}

\subsubsection{Okaramine M: recovering an expert route published after the training cut-off}

Okaramine M (Fig.~\ref{fig:casestudies}a) is proposed to be a key intermediate in the biosynthesis of
the Amauromine class of natural products \cite{schmelzer2025synthesis}, whose members
exhibit, among others, vasodilating activity \cite{takase1984structure} and anticancer
activity \cite{ishikawa2010novoamauromine}. Okaramine M and other Amauromine-class
natural products have been synthesized from the TIPS-protected derivative of Okaramine
M \cite{schmelzer2025synthesis}, demonstrating the synthetic value of this
intermediate, which we selected as our target. 
This example demonstrates that SynthEx can reproduce a route conceived by expert organic chemists and experimentally validated.

The first part of SynthEx's strategy forms the TIPS-monoprotected derivative of
\emph{pre}-Okamauromine through standard protection, condensation and deprotection
reactions. The key step then involves the tandem prenylation of the
\emph{pre}-Okamauromine derivative and formation of the hexahydropyrroloindole core: a
prenyl cation adds at the C3 position of the unprotected indole through electrophilic
addition, followed by trapping of the resulting iminium cation by a nitrogen atom of
the diketopiperazine moiety. In proposing this step, SynthEx displays both a capacity
for elegant chemistry and a grasp of subtle mechanistic considerations: it correctly
identifies the C3 position of the indole as more nucleophilic than C2, and further
recognizes that TIPS protection at the indole nitrogen lowers the nucleophilicity of its
C3, so that electrophilic addition is directed selectively to the C3 of the unprotected indole. 
The synthesis is completed by N-acylation of the indole.
 
Compared with the reference route \cite{schmelzer2025synthesis}, SynthEx follows it
closely: it captures the key tandem prenylation and iminium-trapping step, and it
forms \emph{pre}-Okamauromine through the stepwise condensation of two distinct tryptophan
precursors rather than a direct dimerization of tryptophan, which would give
the diketopiperazine derivative in low yield \cite{schmelzer2025synthesis}.

In one respect SynthEx departs from the published sequence, and the departure may be
an improvement. The literature route installs the TIPS group on
\emph{pre}-Okamauromine, whereas SynthEx protects from the first step, which would
avoid the mixture of mono- and bi-protected \emph{pre}-Okamauromine that late-stage
protection invites. We have not tested this, so it stands as a proposal rather than a
result; but it is a substantive variation on the expert route, not a paraphrase of
it. That SynthEx recovers the strategic logic of an experimentally validated
synthesis published online in June 2025, well after the model's training cut-off, and
offers a defensible refinement of it, illustrates a capacity for retrosynthetic
reasoning rather than recall.

Two limits of this comparison should be stated. The target is the TIPS-protected
intermediate rather than Okaramine M itself, chosen because the Amauromine-class
syntheses proceed through it; and recovery of the strategy was judged by inspection
of the two routes, not by execution. What the case shows is that SynthEx reconstructs
expert strategic logic it cannot have seen, not that its route would perform as
written.

\subsubsection{Melonine: converging on the same key disconnection}

Melonine is a pentacyclic monoterpene indole alkaloid
\cite{kouame2021structure}: an indoline fused to a quinolizidine, carrying the
2,2,3-trisubstituted motif characteristic of the schizozygane alkaloids
\cite{zhang2024vallesamidine} and a two-carbon bridge between the C2 and C20
positions. Its congested, bridged architecture has made Melonine a challenging target, and two total syntheses have been reported to date \cite{matsuyuki2025total, goelo2026total}. Both postdate
the model's training cut-off, Yokoshima's route appearing online in February 2025
and Zhu's route in 2026.

SynthEx's strategy is displayed in Figure~\ref{fig:casestudies}b. It begins by iodination
of Boc-protected tryptamine followed by 
deprotection. The next steps involve a reductive amination and an intramolecular Heck
coupling. The key step is a highly elegant tandem reaction: the condensation of
the secondary amine in the substrate and formaldehyde enables an aza-Cope rearrangement
to take place; the newly formed iminium undergoes a Mannich reaction with the nucleophilic
C3 position of the indole, followed by a migration of an alkyl chain from the C3 to the 
C2 position, these last two steps mimicking a Pictet--Spengler reaction; the cascade ends
with a reduction, unlike a classical Pictet--Spengler that would restore aromaticity. 
It is to note that a cyclization through Mannich reaction is
also proposed in the biosynthesis of Melonine, starting from a reduced form of
the intermediate that SynthEx employs \cite{kouame2021structure}.
Finally, SynthEx reaches the target Melonine by selective hydrogenation.

SynthEx's strategy differs from Melonine's two published total syntheses: Yokoshima and co-workers have as key steps an oxidative intramolecular aziridination followed by nucleophilic attack
of an arylamine \cite{matsuyuki2025total} while Zhu and co-workers
employ as key steps a bis-cyclisative diamination
followed by a lactamization \cite{goelo2026total}.
Zhu and co-workers moreover attempted a route along the proposed biosynthetic
pathway, centered on exactly the Mannich cyclization that SynthEx selects as its key
step. That attempt failed, and the reason is conformational: in their substrate the
conformation required for cyclization suffers a severe steric clash between the
piperidine ring and the C--H bonds of the CH$_2$CH$_2$ linker, so the iminium cannot
adopt a geometry from which the indole can reach it. Full experimental detail is in
the Supporting Information of the work \cite{goelo2026total}.

One of us (J.Z.) led those experiments, which allows the comparison with SynthEx's
route to be made directly rather than inferred. The distinction is the linker. Because
SynthEx generates its iminium through an aza-Cope rearrangement rather than by
condensation, the CH$_2$--CH$_2$ single bond of the failed substrate is replaced by an
HC$=$CH double bond. That change removes the clash, and the resulting intermediate
should reach a reactive conformation considerably more easily. 
A similar cyclization on a related substrate with similar reactive conformation has been realized by the same group \cite{delayre2020ticl3}. 
On present evidence, SynthEx's route is therefore a live
experimental proposal rather than a repeat of a known failure, and it is one we
intend to test.

The convergence is worth separating from the outcome. Given only the structure, and
with no published route to Melonine available to it, SynthEx identified the same
disconnection that an expert group judged worth committing laboratory effort to.
Agreement with an idea experts chose to test is a demanding standard because it is
independent of whether the idea worked. That SynthEx then reached the same
disconnection by a route which, on the assessment of the group that ran the original
experiments, may succeed where theirs did not, is a stronger result than convergence
alone.

\subsubsection{Chanoclavine to Lysergol: planning from an advanced intermediate}

A synthetic campaign does not always begin with simple commercial materials. A group may hold a
hard-won advanced intermediate and be unable to close the remaining gap to the
target, in which case the useful question is not how to make the molecule but how to cross the finishing line from what is already accomplished. SynthEx accepts a required starting material
as a constraint on strategy generation (Section~\ref{res:agent},
Methods~\ref{methods:synthex}), which turns it on that problem directly. We tested
this on the unprecedented conversion of Chanoclavine, a tricyclic ergot alkaloid that has been produced on scale through a combination of synthesis and bioengineering \citep{ma2022hybrid}, into Lysergol,
which requires forming the D ring of the ergoline skeleton
(Fig.~\ref{fig:casestudies}c).

Given only the two structures of Chanoclavine and Lysergol, the \textit{Strategy Generator} returned a strategy centered on a Hofmann--L\"{o}ffler--Freytag
reaction: chlorinate the secondary amine, photolyze to generate the nitrogen radical,
abstract a hydrogen through a 1,6-hydrogen atom transfer (1,6-HAT) to functionalize an
allylic methyl group, and use the resulting alkyl chloride as the
electrophile for a base-promoted intramolecular N-alkylation that closes the D ring,
forming Elymoclavine, which upon a double-bond migration, is converted into Lysergol.
Even though a 1,6-HAT is far less common than a 1,5-HAT, such a reaction has literature
precedence \citep{dupeyre1973application}. Moreover, the rigidity of the substrate,
the absence of any H atom correctly aligned for a 1,5-HAT, and the stability of 
the allylic radical that would be formed through 1,6-HAT are all factors that
support the feasibility of this step in SynthEx's strategy. It is also noteworthy to mention
that SynthEx's strategy is a unique way to reach Lysergol from
Chanoclavine, differing from the proposed biosynthetic pathway whose key steps
involve the oxidation of Chanoclavine to Chanoclavine-aldehyde followed by 
condensation to reach an iminium cation \cite{jakubczyk2014biosynthesis}.

Where the conventional disconnection for this ring closes onto a carbon that already bears a functional handle, SynthEx instead installs the handle 
where the chemistry needs it, through remote functionalization of an unactivated C–H bond. This is precisely the kind of low-frequency synthetic
tactics that Section~\ref{res:space} shows to be scarce in the patent record, and SynthEx proposed it unprompted to bridge a specific two-compound
gap.

This is the mode of use we expect to matter most in practice. A campaign stalled a
few steps from its target has already heavily invested in the formation of its advanced intermediates, and the
value of a planner there is not a route from commercial material but alternative ways
to succeed in the last few steps. Because strategies can be conditioned on an
intermediate, SynthEx can be asked that question directly.

The three targets were selected for their utility to showcase the capability of the pipeline, but we note that this level of chemical sophistication is not isolated. Experts engaged with the proposed steps as chemistry to be reasoned about rather than as output to be scored, a response that algorithmic route prediction has not previously drawn. We now move on to quantitative evidence to examine whether this qualitative distinctiveness holds systematically.

\subsection{Analysis of SynthEx's reaction space}
\label{res:space}
Because the model writes disconnections rather than choosing them from a library (Section~\ref{res:agent}), SynthEx is not bound by the fixed templates and patent-frequency priors of conventional planners. We hypothesized that this would push it into regions of reaction space that patent-trained tools systematically under-represent; that the resulting chemistry would be practically useful rather than merely unusual; and that much of it would be absent from the proposals of even a state-of-the-art single-step model. We tested each prediction
in turn, and then asked which chemistry accounts for the difference.

\begin{figure*}[h!]
  \centering
  \includegraphics[width=\textwidth]{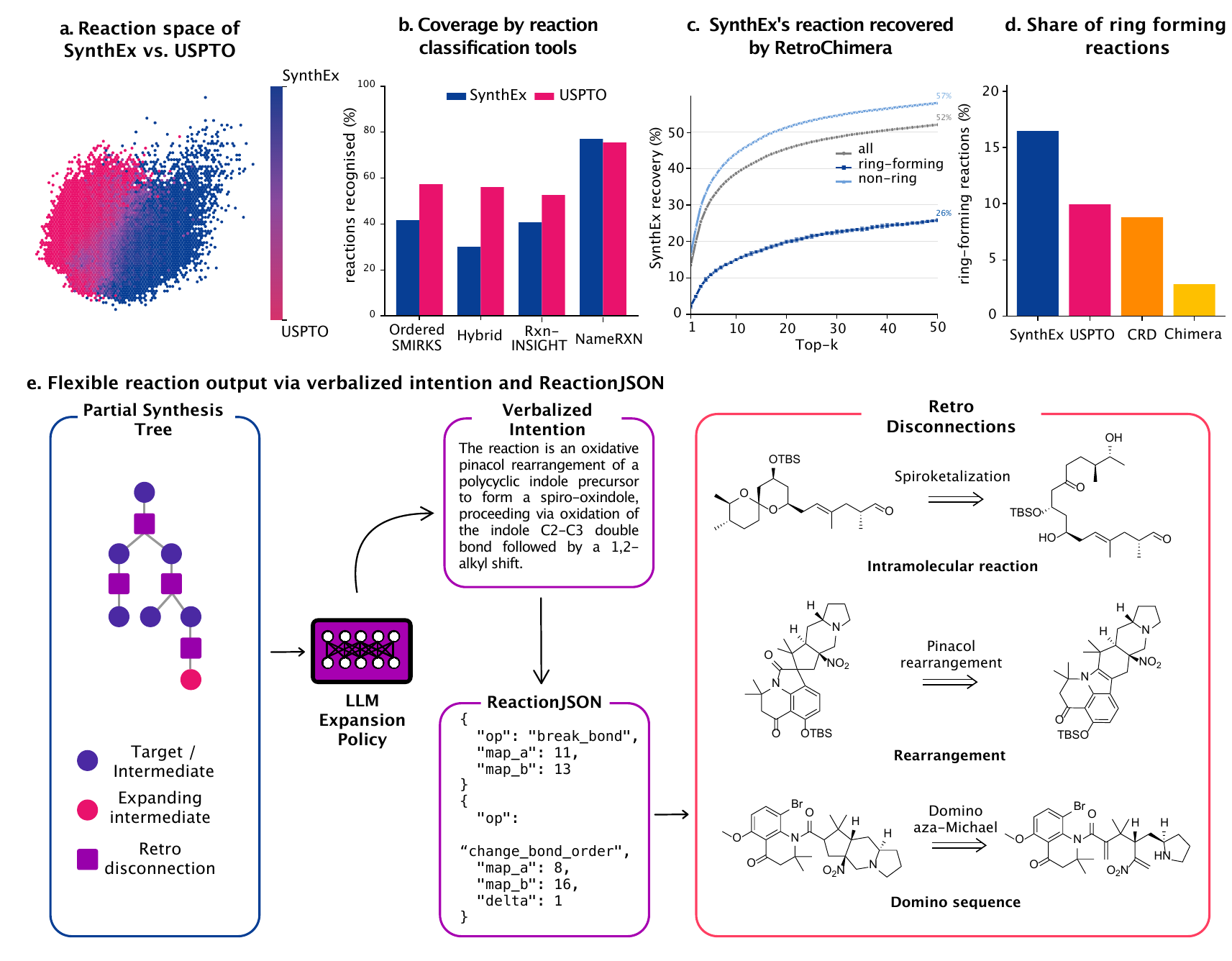}
  \vspace{1mm}
  \caption{\textbf{SynthEx occupies a distinct reaction space that template-based
  tools cannot reproduce, enriched in ring construction.} \textbf{a}, ~PCA of the output layer of a neural classifier co-embedding of SynthEx reactions (blue) with USPTO reactions (pink);
  SynthEx reactions form a largely contiguous territory distinct from the bulk of
  the patent-derived taxonomy. \textbf{b}, ~Fraction of reactions recognized by
  four classifier configurations for SynthEx versus a random USPTO sample; the two
  corpus-derived methods (ReactionClassifier, Ordered and Hybrid) recognize
  15--25~percentage points fewer SynthEx reactions, whereas the
  corpus-independent NameRXN dictionary is at parity. \textbf{c}, ~Fraction of
  SynthEx ground-truth disconnections recovered within RetroChimera's top-$k$
  predictions ($k=1$--50) for all steps, ring-forming steps and non-ring-forming
  steps; recovery of ring-forming disconnections saturates far below completeness.
  \textbf{d}, ~Ring-forming reactions as a fraction of each corpus; SynthEx
  (16.0\%) far exceeds USPTO, recent academic reactions (CRD), and RetroChimera's
  top-1 predictions (2.8\%). \textbf{e}, The \textit{Route Builder}'s
  natural-language reasoning, its ReactionJSON graph-edit output, and
  representative ring constructions.}
  \label{fig:space}
\end{figure*}

We first characterize the SynthAtlas reaction corpus of 33,145 steps produced by running SynthEx across the more than 1,000 natural-product targets (Fig.~\ref{fig:pipeline}c, Section~\ref{methods:curation}), by how readily existing tools recognize it.
We use three classification tools, one of them in two modes: NameRXN, Rxn-INSIGHT
and ReactionClassifier \citep{namerxn,rxn-insight,armstrong2026agentic}. ReactionClassifier is a hierarchical neuro-symbolic namer in which a
fingerprint-based Multi Layer Perceptron proposes a class within a corpus-derived taxonomy
and the label is returned only if one of the retrosynthetic templates belonging
to that class reproduces the recorded product, so that recognition requires an
explicit template match.
The decisive comparison is NameRXN, an expert-curated dictionary tied to no
training corpus, which is essentially at parity between SynthEx and USPTO
(Fig.~\ref{fig:space}b): the reactions are therefore not ill-defined or
unnameable. What separates them is their origin rather than their validity. The
USPTO-derived ReactionClassifier recognizes 15--25 percentage points fewer SynthEx
reactions than USPTO reactions, confirming that this nameable chemistry is scarce in
the patent record. 

The two corpora also separate geometrically, a principal-component projection of ReactionClassifier's neural network's output layer
places them in largely distinct regions rather than dispersing them through one
another (Fig.~\ref{fig:space}a); we show this as a visualization rather than as
independent evidence, since that classifier is itself trained on patent reactions and
some separation follows from that alone.
 
For a template-based tool, non-recognition directly implies non-reproduction, as a reaction absent from the library cannot be proposed. Models that construct disconnections rather than retrieving them from a library are not bound in this way. Because they predict precursors directly, either by editing the product graph or
by generating them \textit{de novo}, they can in principle express any
transformation, yet they remain anchored to the distribution of its training reactions, which is again the patent corpus; representational freedom is not distributional freedom. We therefore tested reachability against RetroChimera~\cite{maziarz2024chimera}, a state-of-the-art single-step model that ensembles a graph-editing component and a \textit{de novo} Transformer with complementary inductive biases, and which its authors report to remain robust outside its training data. Presenting the product of each SynthEx reaction to RetroChimera (trained on Pistachio), the SynthEx disconnection appears in its top-1 prediction for only 13.5\% of steps and its top-5 for 31.4\% (Fig.~\ref{fig:space}c); therefore, more than two-thirds of SynthEx's transformations are absent from the top-5 of a leading corpus-trained model.
This gap is most pronounced in SynthEx's signature chemistry—ring-forming disconnections—where top-1 recovery falls to 2.3\% and top-5 to 10.9\%.
Deeper prediction lists do not lift the ceiling, RetroChimera recovering only 52.0\%
of disconnections overall and 25.8\% of the ring-forming ones even at top-50. Top-50
is in any case a generous regime: a multi-step search at the depths these targets
demand can expand only a handful of candidates per node, so a disconnection buried far down a ranked list is rarely reached in practice. The chemistry SynthEx proposes is thus not a faster path to the same routes, but a region that a template library cannot reproduce and that a state-of-the-art corpus-trained model rarely proposes.
 
Finally, we examine what chemistry drives this separation. The single clearest
signature is ring construction (Fig.~\ref{fig:space}d): SynthEx forms a ring in
16.0\% of its steps, against 9.9\% for USPTO, 8.7\% for recent academic
reactions, and only 2.8\% for RetroChimera's own top-1 predictions. Ring formation
is precisely the low-prior, high-entropy chemistry described above: a large,
heterogeneous family of individually rare transforms, poorly served by rigid traditional retrosynthesis models. Whereas NameRXN recognizes two-thirds of SynthEx’s ring-forming steps, USPTO-derived methods identify fewer than one in ten (Table~\ref{tab:rings}), indicating how poorly such reactions are served by USPTO-derived tooling.
 
This ring-building bias is one facet of a broader constructive signature. C--C bond
formation is the single largest class of SynthEx's named steps, at 22.5\%, more
than double its share of RetroChimera's predictions on the same targets (9.7\%);
and these constructions are predominantly convergent, with 63.5\% uniting two
independent fragments through aldol addition, Horner--Wadsworth--Emmons and
Julia--Kocienski olefination, Suzuki and Stille coupling, or olefin metathesis, and
the remaining 36.5\% closing rings intramolecularly. The corpus-trained model
instead defaults to conservative functional-group editing: protecting-group
manipulations account for 40.0\% of RetroChimera's top-1 disconnections but only
27.0\% of SynthEx's steps, and RetroChimera strongly over-proposes reduction and oxidation reactions
and protection adjustments while under-proposing organometallic coupling. 

Furthermore, ReactionJSON lets the language model express many different modes of ring construction and reorganization. Figure~\ref{fig:space}e illustrates this range with three examples: an intramolecular spiroketalization, a pinacol rearrangement of a polycyclic skeleton, and a domino aza-Michael that closes several rings in one operation.

\begin{table}[t]
  \centering
  \caption{\textbf{Recognition of SynthEx's ring-forming steps.} Fraction of the
  5,318 ring-forming and 27,827 non-ring-forming SynthEx steps named by
  each classifier. NameRXN, an expert-curated dictionary independent of any
  training corpus, names two-thirds of the ring-forming steps; every
  USPTO-frequency-derived method recognizes almost none. ReactionClassifier
  (Ordered) and ReactionClassifier (Hybrid) are the two variants of our
  template-based classifier (Section~\ref{res:space}). The ratio is ring-forming
  over non-ring-forming recognition; a value near unity indicates no ring-specific
  deficit.}
  \label{tab:rings}
  \begin{tabular}{lccc}
    \toprule
    Classifier & Ring-forming & Non-ring-forming & Ratio \\
    \midrule
    NameRXN                        & 66.8\% & 79.1\% & 0.84 \\
    Rxn-INSIGHT                    & 3.2\%  & 47.8\% & 0.07 \\
    \addlinespace
    ReactionClassifier (Ordered)   & 6.9\%  & 48.4\% & 0.14 \\
    ReactionClassifier (Hybrid)    & 2.2\%  & 35.6\% & 0.06 \\
    \bottomrule
  \end{tabular}
\end{table}

The chemistry demonstrated in the Okaramine, Melonine and Lysergol routes, ring-forming, convergent, and drawn from named-reaction rather than patent-frequency space, is not a property of those three targets; it is representative of the key steps found in the SynthAtlas corpus.

\subsection{Reach across the benchmark and expert assessment of the chemistry}
\label{res:reach}

The above analysis has confirmed why SynthEx succeeds where traditional tools fail, but does not establish how often SynthEx succeeds — on what fraction of complex targets it returns a complete route to purchasable building blocks. We quantify that reach across the benchmark next.
Existing multistep retrosynthesis benchmarks based on the patent literature, such
as USPTO-190 \citep{chen2020retro}, PaRoutes \citep{genheden_bjerrum_2022} and Pistachio-Hard \citep{yu2024double}, are by and large saturated, with
state-of-the-art tools suggesting reasonable routes within a sensible compute
budget \citep{genheden2020aizynthfinder,chen2020retro, pdvn, maziarz2024chimera}.
However, such benchmarks are largely in distribution for existing models, and do
not permit claims about performance on the frontier of organic chemistry, the total
synthesis of natural products. We therefore benchmark SynthEx on 1,098 structurally complex natural products drawn from NP-Atlas (Methods~\ref{methods:curation}). A target is scored as solved only when a complete route to purchasable building blocks is found.
 
For a conventional planner, these targets are effectively out of reach even under a generous compute budget. 
We use AiZynthFinder as the baseline because it is a widely adopted open-source template planner and, being inexpensive per expansion, can be run near-exhaustively; its failures therefore reflect the limits of the patented reaction space rather than an exhausted search budget.
Run near-exhaustively over our building-block stock,
without a practical expansion cap, to a maximum search depth of 25 and a wall-clock limit
of 30 minutes per target, AiZynthFinder expands a median of roughly $2.9\times10^{4}$
nodes yet solves only 13.8\% of the benchmark (151/1,098; Methods~\ref{methods:search}).
The barrier is therefore not search budget but reaction space: the disconnections
these targets require lie outside the template library, so no amount of additional
expansion within it reaches them.

\begin{figure*}[t]
  \centering
  \includegraphics[width=\textwidth]{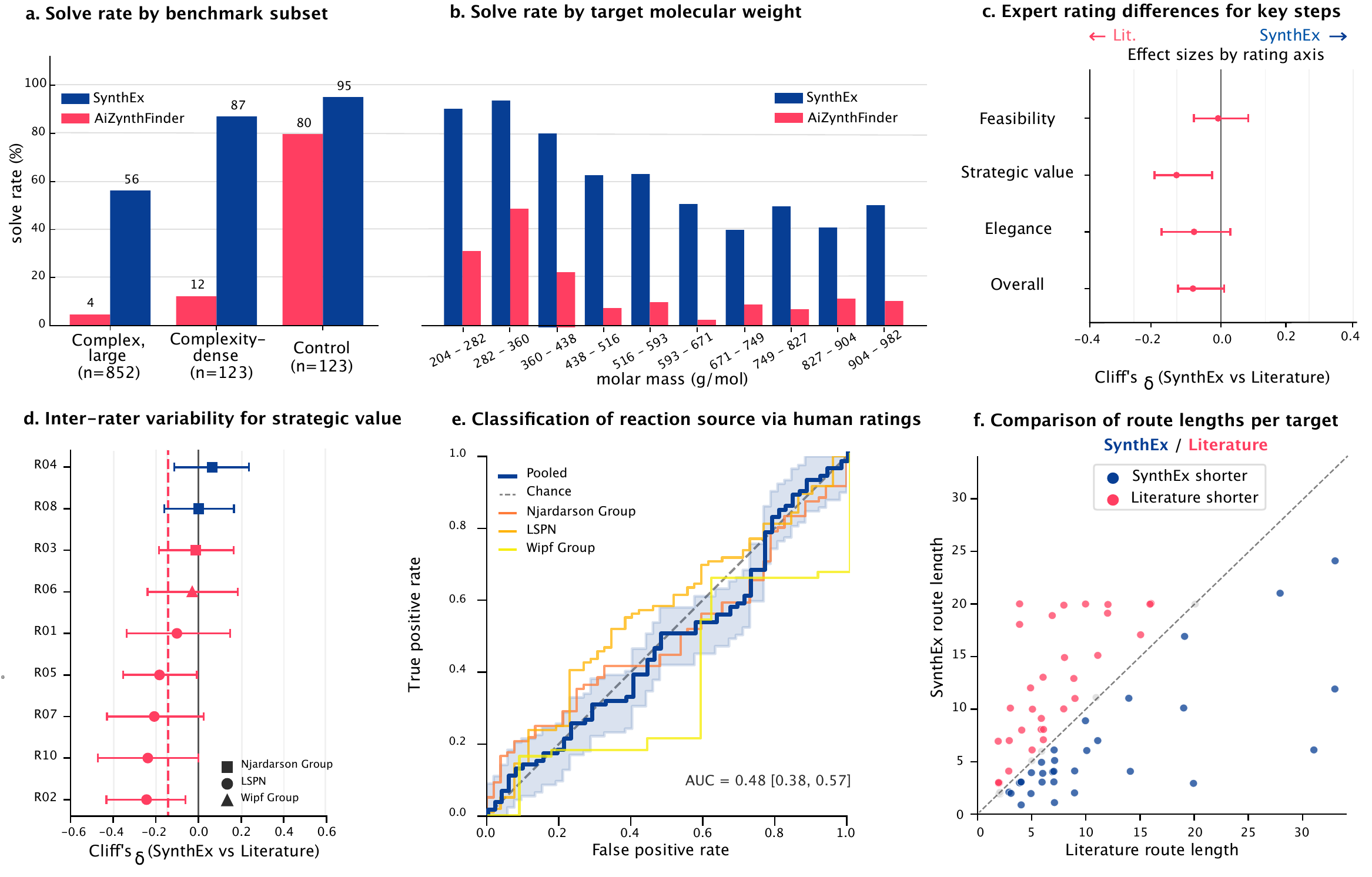}
  \vspace{1mm}
  \caption{\textbf{General results on the SynthEx benchmark.} \textbf{a}, Comparative solve rates of SynthEx and AiZynthFinder on several benchmark subsets, detailed in Supplementary Information: Benchmark Construction. \textbf{b}, Performance degradation relative to target molecular weight in Daltons. \textbf{c}, Differences in expert ratings for key steps (SynthEx vs. literature) across feasibility, strategic value, elegance, and overall axes. \textbf{d}, Inter-rater variability (heterogeneity) on the strategic value axis. \textbf{e}, Predictive performance of human ratings, displaying the AUC of a logistic classifier used to predict whether a key step originates from the literature or SynthEx. \textbf{f}, Target-wise comparison of route lengths between academic literature and SynthEx routes.}
  \label{fig:results}
\end{figure*}

The benchmark subsets make the same point in a way that also rules out a
misconfigured baseline (Fig.~\ref{fig:results}a). On the control subset, which was
assembled from structurally undemanding targets, AiZynthFinder solves 80\%, close to the performance such tools report on the patent-derived
benchmarks they were built for; SynthEx solves 95\%. On the more complex targets, the performance of AiZynthFinder drops, to
12\% on the complexity-dense set and only 4\% on the large
complex set, while SynthEx retains most of its reach. The failure we report is
specific to structural complexity rather than a property of how the baseline was
configured.

The control subset sharpens this into something close to a controlled comparison,
because of how it was selected. Those targets were chosen precisely because a short,
resource-limited template search had failed on them
(Methods~\ref{methods:curation}); given the generous budget used here, the same
planner solves 80\% of them. Their original failure was therefore a budget failure,
and more compute repairs it. The complex subsets were run under exactly the same
generous budget and are not repaired: the planner still returns nothing for 88--96\%
of them. The same intervention that rescues one group leaves the other untouched,
which is the distinction between a search that is under-resourced and one that is
looking in the wrong space.
 
SynthEx overcomes this barrier by widening the reaction space the search can draw on, not by searching the existing one harder. By guiding toward articulated key steps, its language-model strategist proposes complexity-reducing
disconnections that the template library does not contain, simplifying each target
toward precursors a short template search can finish; unsolved leaves are then
completed by an AiZynthFinder call under a deliberately small budget (maximum depth
6), which succeeds precisely because the leaves handed to it are simple. The
effect is cumulative (Table~\ref{tab:solve_rate}): the strategic layer alone,
without any leaf completion, already solves 25.0\% of targets (275/1,098),
exceeding the exhaustive template baseline; adding the short-budget completion of
its leaves raises the target-level solve rate to 63.9\% (702/1,098). The same
tool that fails as a stand-alone planner succeeds as a completion
engine once SynthEx has done the strategic work of substantially decomplexifying the target product through elegant, constructed steps, a division of labor whose
advantage widens sharply with molecular complexity (Fig.~\ref{fig:results}b). 

\begin{table}[t]
  \centering
  \caption{\textbf{Reach on the natural-product benchmark ($n$~=~1,098).}
  AiZynthFinder is run near-exhaustively (no expansion cap, maximum depth 25,
  30 minutes per target); SynthEx's strategic layer is the language-model
  decomposition alone; the stitched result adds a short-budget (maximum depth 6)
  AiZynthFinder completion of the remaining unsolved leaves. The same template
  engine that solves only 13.8\% as a stand-alone planner completes SynthEx's
  simplified leaves to a 63.9\% target-level solve rate.}
  \label{tab:solve_rate}
  \begin{tabular}{lcc}
    \toprule
    Method & Solved & Solve rate \\
    \midrule
    AiZynthFinder (exhaustive baseline)        & 151 & 13.8\% \\
    SynthEx, strategic layer only              & 275 & 25.0\% \\
    SynthEx, stitched (strategy + completion)  & 702 & 63.9\% \\
    \bottomrule
  \end{tabular}
\end{table}
 
Solve rate alone is an inadequate measure of a route's quality, with the feasibility
of the individual reactions mattering at least as much
\citep{segler2018planning, maziarz2023re}. We therefore asked expert chemists a
narrower and more answerable question than ``is this a good route'': holding the
overall strategy fixed, how do the reactions SynthEx selects compare with those a
chemist selected for the same target?

Conditioning on a shared strategic frame is what makes the comparison interpretable,
since it isolates the quality of the chemistry from differences in overall plan. We
assembled 70 natural-product targets, each with a total synthesis published after the
model's training cut-off and for which SynthEx also returned a route to the same
target, and retained the 47 on which SynthEx independently arrived at a strategy
congruent with the published one. Key steps were drawn from both routes, rendered
identically and stripped of any cue to their origin, and rated by ten synthetic
chemists from total-synthesis groups on four axes: feasibility, strategic value,
elegance and overall quality. This yielded 1,040 ratings over 148 unique key steps
(Methods~\ref{methods:rating}).

Individual raters are not independent replicates; each rates many items,
and the items themselves are rated by an uneven number of raters. Hence we treat the
rater, not the rating, as the unit of resampling: all intervals below come from a
cluster bootstrap over the ten raters (resampling raters with replacement and
recomputing item means from whichever ratings survive), which is the appropriate
correction and is applied throughout rather than chosen post hoc. On this basis,
SynthEx key steps were statistically indistinguishable from the published expert
steps on feasibility ($\delta = -0.01$, 95\% CI [$-0.09$, $+0.08$]), elegance
($\delta = -0.09$, [$-0.19$, $+0.03$]) and overall quality ($\delta = -0.09$,
[$-0.14$, $+0.01$]), where $\delta$ is Cliff's delta over per-item mean scores and
positive values favor SynthEx (Fig.~\ref{fig:results}c). Strategic value is the one
axis with a detectable gap ($\delta = -0.14$, [$-0.21$, $-0.03$]), which remains on
the literature-favoring side even under Holm correction for testing four axes
(98.75\% CI [$-0.24$, $-0.01$]). Equivalently, the largest true difference consistent
with the data is small on every axis: $|\delta|\le0.08$ for feasibility, $0.20$ for
strategic value, $0.18$ for elegance and $0.13$ for overall.

This one detectable gap is smaller than the disagreement among the raters themselves.
Recomputing $\delta$ from each rater's own scores alone, the standard deviation
across the ten individual estimates exceeds the pooled effect on every axis,
including strategic value (ratio $1.47$; elegance $1.96$; overall $1.10$;
Fig.~\ref{fig:results}d), and it is not shared evenly: one participating group shows
a clear literature preference on this axis while the other two show essentially
none. We directly tested whether the four-axis rating profile can identify a step's source: a logistic classifier on the per-item mean scores, evaluated by leave-one-item-out cross-validation, gives an area under the ROC curve of $0.48$ (95\% CI [$0.38$, $0.57$]), indistinguishable from a permutation null (Fig.~\ref{fig:results}e); the panel's own scores carry no detectable signal about which steps are SynthEx's. Strategic value is therefore the one axis worth reporting as a real, if small and rater-dependent, difference; with strategy controlled, whatever separates expert from machine on that axis lies in higher-order planning rather than in the soundness of the steps themselves, and it is not a difference the expert panel could reliably act on step-by-step.

That is also the one stage at which SynthEx is built to accept human direction. The
\textit{Strategy Generator} takes chemist-supplied strategies and constraints in natural
language (Section~\ref{res:agent}), and we withheld that input throughout in order to
measure the system unaided. However, whether such steering
would narrow the gap is a question these data cannot answer.

On targets where both a literature synthesis and a SynthEx route exist,
SynthEx routes are frequently the longer of the two (Fig.~\ref{fig:results}f, right).
This is expected rather than damaging. A published total synthesis is the endpoint of
an optimization campaign in which steps were telescoped, protecting groups designed
out and sequences shortened over months of laboratory work, whereas a SynthEx route
is a first proposal that has never met an experiment. The comparison worth drawing is
with what an automated planner produces, not with the distilled result of a human
campaign; brevity relative to the literature is a target for the field, not a claim
we make here. In the SI \ref{si:routelen} we provide a similar figure comparing route lengths where both AiZynthFinder and SynthEx find a route, with SynthEx consistently finding shorter pathways for a given target.

\subsection{Iterative and granular synthetic route improvement}
\label{res:improve}

Iterative feedback loops have proven effective in coding agents
\citep{shinn2023reflexion,takerngsaksiri2025human}, and the case for one is stronger
still in retrosynthesis, where a proposed step can fail in more ways than a line of
code can. The disanalogy matters as much as the parallel. A coding agent is corrected
by a compiler and a test suite, which are ground truth; synthesis planning has no
such oracle short of the laboratory. We therefore use a language-model critic as a
stand-in, and are explicit below about what that does and does not establish.

\begin{figure}[h!]
    \centering
    \includegraphics[width=0.95\linewidth]{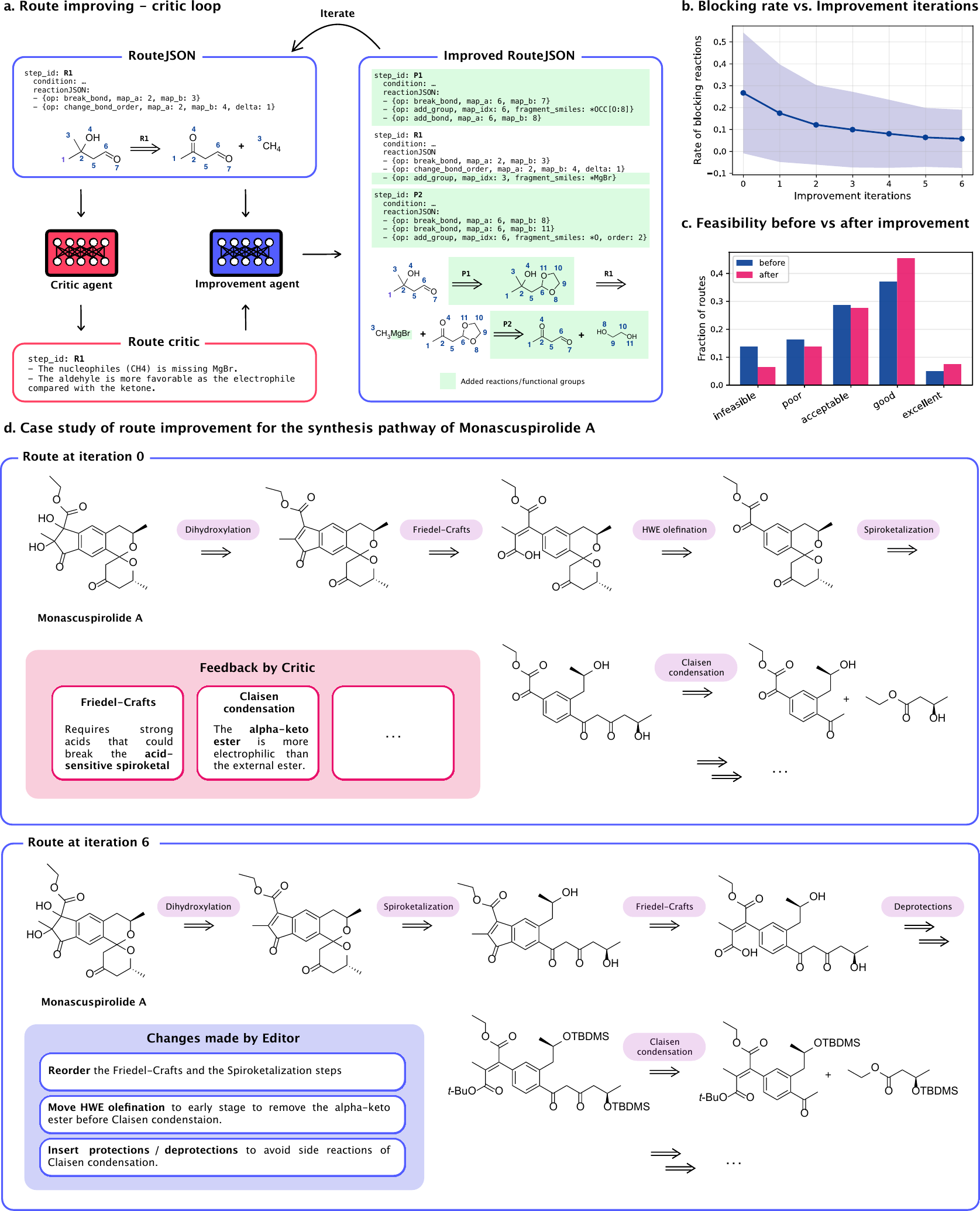}
    \vspace{1mm}
    \caption{\textbf{SynthEx's iterative route refinement.} \textbf{a}, The iterative action--feedback loop between the \textit{Critic} and the \textit{Editor}. \textbf{b}, The average per-route blocking-reaction rate decreases as the number of iterations increases. \textbf{c}, Overall route feasibility across SynthAtlas before and after the improvement loop, showing fewer \textit{infeasible} and \textit{poor} routes and more \textit{good} and \textit{excellent} ones. \textbf{d}, A worked example of the pipeline applied to a synthetic route for Monascuspirolide A.}
    \label{fig:route_improve}
\end{figure}

Inspired by how contemporary coding agents interact with source files through surgical editing and exact text replacement \cite{yang2024swe}, we translate each synthetic route into a RouteJSON document, a linear sequence of ReactionJSON entries (Fig. \ref{fig:route_improve}a). ReactionJSON anchors its graph-edit operations on atom maps, thereby avoiding branching \cite{xuan2025tempre,wang2025llm} and allowing the agent to focus on the transformation itself rather than rewriting whole molecules \cite{edwards2022translation,silly_things_walters,jang2024llmsgeneratediversemolecules}.
This enables flexible manipulation of the synthetic tree through the textual modality, extending beyond the insertion of protections and deprotections \cite{westerlund2026toward}: the agent can reorder reactions, insert or delete reaction steps, and alter reaction conditions or functional groups.

Before surfacing a route to human users, the third phase of SynthEx runs an iterative action--feedback loop between a \textit{Critic} and an \textit{Editor} (Fig. \ref{fig:route_improve}a). The \textit{Critic} acts as a reaction harness, simulating each reaction in the forward direction and flagging chemically infeasible steps, termed \textit{blocking} reactions \cite{bran2026chemical}. The \textit{Editor} is then tasked with resolving all blocking reactions through surgical route edits while keeping the core strategy of the route intact. 
The resulting route is then re-evaluated by the \textit{Critic} agent, and the loop continues until either all blocking reactions are resolved or a maximum number of iterations is reached.

The blocking-reaction rate, computed per route as the number of blocking reactions
divided by the total number of reactions, falls steadily with iteration, from
approximately 0.27 before any repair to approximately 0.06 after six iterations
(Fig. \ref{fig:route_improve}b).
The feasibility distribution shifts
accordingly when the finished routes are re-scored by the \textit{Analyst} in phase 4,
with fewer \textit{infeasible} and \textit{poor} routes and more \textit{good} and
\textit{excellent} ones (Fig. \ref{fig:route_improve}c).

We note that this feasibility criterion rests on the knowledge of the language models themselves. While they can flag many obvious problems, certain blind spots may be shared across the agents, since they share the same LLM backbone. Establishing true feasibility requires experimental validation.

Figure \ref{fig:route_improve}d illustrates this pipeline on a concrete SynthAtlas target, Monascuspirolide A, whose route quality improves markedly after the improvement loop.
The original route proposed by the \textit{Route Builder} in phase 2 is flagged with several feasibility issues. First, the acid-labile spiroketal ring is installed mid-route, preventing subsequent steps from being run under acidic conditions; indeed, as the \textit{Critic} agent notes, the late-stage Friedel--Crafts reaction, typically conducted with a Brønsted or Lewis acid, could decompose the ring.
To resolve this blocking step, the \textit{Editor} relocates the spiroketalization to the final step, after the Friedel--Crafts reaction, minimizing the ring's exposure across the synthesis.
Second, in the Claisen condensation step, the \textit{Critic} agent observes that the acetophenone substrate carries an $\alpha$-keto ester group that is more electrophilic than the ester of the external substrate, potentially leading to polymerization.
The \textit{Editor} resolves this by moving the Horner--Wadsworth--Emmons (HWE) olefination to an earlier stage, eliminating the ketone at the carbon $\alpha$ to the ester group of the acetophenone.
Finally, the acid introduced after HWE olefination is protected as its \textit{tert}-butyl ester and the alcohols are protected as TBDMS ethers, shielding these sites from the strong base used in the Claisen condensation.

\subsection{SynthAtlas: an open resource over natural-product synthesis}
\label{res:synthatlas}

As a result of our benchmark, we release \textbf{SynthAtlas}, an open resource of SynthEx-designed syntheses of natural products (Fig.~\ref{fig:pipeline}c). The release comprises 1,098
natural-product targets, 3,243 routes with synthetic strategies (a mean of 2.95 per
target) and 33,145 fully specified reaction steps (mean 10.22 steps per route).
Some strategies and corresponding key reactions are illustrated in Fig.~\ref{fig:pipeline}d.
Every reaction carries a complete atom-mapping, produced by construction from the
graph-edit representation rather than by a post hoc mapper, together with structural
descriptors including ring-formation flags and reactant count. Because the mapping
falls out of the representation, the corpus is internally consistent in a way that
post hoc mapped datasets are not.

Routes are browsable through an interactive platform, on which each strategy can be
inspected step-by-step alongside the agent's reasoning, compared against the
alternative strategies proposed for the same target, and commented on by other
chemists.

The resource is intended to be useful in two ways. Because it captures the
convergent, ring-building chemistry that current retrosynthesis models
under-represent (Fig.~\ref{fig:space} and Table \ref{tab:rings}), and because its
individual steps were rated comparably with published expert steps
(Section~\ref{res:reach}), it offers training and evaluation material of a kind the
patent corpora do not supply. And because every target was selected as having no
reported total synthesis at the time of curation, each released route is a dated,
public prediction about a molecule nobody has yet made. We intend to report
concordance as syntheses of these targets appear.

\section{Discussion}\label{sec13}

The multistep benchmarks the field has long used are now highly saturated.
Patent-derived planners report success approaching completeness on them, so
differences in score no longer resolve differences in capability, much as early
coding benchmarks stopped separating models before the field moved from isolated
scripts to repository-scale software engineering tasks
\citep{jimenez2024swe,yang2024swe}. Solve rate has not stopped being a meaningful
quantity; it has stopped varying on those targets. We therefore evaluated on the
regime where the open problem lies, a set enriched for the structural complexity
that defeats standard planning tools and against which a near-exhaustively resourced
template planner solves only 13.8\%
\citep{maziarz2023re,tripp2024retro,morgunov2026syntax}. On this set SynthEx returns
complete routes for 63.9\% of targets, close to a five-fold increase in reach, and
its margin widens rather than narrows as the molecules grow more complex. Reach matters
most exactly where existing tools reach least: a planner that returns a route for
one target in eight leaves the great majority of the frontier untouched, however
sound its chemistry may be for the remainder.

What makes this possible is a change in the chemistry available to the planner, not
merely in the efficiency of the search. SynthEx plans in a region of reaction space
that catalogue-based planners cannot reproduce and corpus-trained models rarely
propose, favoring the convergent, ring-forming, bond-constructive steps that
characterize expert total synthesis and that the patent record systematically
under-represents. On individual targets, this chemistry withstands inspection. For
Okaramine M it recovers the strategic logic of an expert route published after the
cut-off and offers a defensible refinement of the protecting-group sequence; for
Melonine it converges, unaided, on the same difficult disconnection an expert group
judged worth attempting, and reaches it by a variant that the group who ran those
experiments assess as more likely to succeed than their own. In blinded assessment, the differences
between its key steps and published expert steps were small on all four axes, with strategic value the remaining edge held by literature chemistry.

We would go further and propose complex natural-product synthesis as the setting in
which planners should now be measured. It probes chemical reasoning more directly
than any patent-derived set, since the disconnections it demands are by construction
absent from the catalogues, and its difficulty scales continuously with structural
complexity rather than saturating. Capabilities of this kind tend to look
discontinuous from the outside, a task on which every system scores near zero
admitting one that scores ten per cent and ceasing to separate systems at all not
long after; a benchmark is most useful before that happens, and natural-product
planning is still early on that curve.

The routes are the second output. SynthAtlas releases 1,098 targets, 3,243 routes with strategies
and 33,145 atom-mapped steps, a body of convergent, ring-forming chemistry of a kind
the patent corpora do not contain and against which the next generation of planners
can be trained and evaluated. Because every target was selected as having no reported
total synthesis, the release is also a set of dated, public predictions about
molecules nobody has yet made, and we will report concordance as syntheses of them
appear.

What we report is reach and per-step quality, not experimental feasibility. We have
not verified stereochemical outcomes, and expert review surfaced occasional
selectivity and feasibility errors; the expert comparison is conditional on a shared
strategic frame, so it speaks to the chemistry SynthEx proposes once a viable
strategy is found rather than to how reliably it finds one; the improvement loop is
scored by the same class of model that performs the repairs, so it measures internal
convergence rather than experimental validity; and the language-model backbone remains
costly relative to a template search. A route a chemist judges sound on paper is a
hypothetical evaluation, not a result.

The axis on which expert judgment still holds an advantage is also the most amenable to user provision through natural-language strategy descriptions. The most productive near-term arrangement may therefore be neither autonomous planning
nor unaided human design, but a chemist supplying the strategy and the agent working
it out. Beyond that, the integration of reaction conditions and, ultimately,
closed-loop validation in the laboratory remains the next frontier, and the planners
that follow should be judged against experimental proof, on targets where the question
is still open.
\backmatter
\section{Methods}\label{sec:methods}

\subsection{Curation of natural-product targets}\label{methods:curation}

The benchmark was drawn from the NPAtlas database (release 2024\_09), retaining only natural products with no reported total synthesis according to NPAtlas.
From these we selected structurally complex molecules, using a Bertz complexity of 900–2200, 24–65 heavy atoms, 4–12 stereocenters, 2–8 rings and at most 14 rotatable bonds. We then clustered them by Morgan-fingerprint similarity with BitBirch~\cite{perez2025bitbirch}, keeping the medoid of each cluster and retaining sparse clusters so as to span the chemical space of the collection. This representative core forms the complex, large subset (n = 852).

We further added two smaller subsets chosen to probe specific failure modes. The complexity-dense subset (n = 123) comprises molecules for which AiZynthFinder's route is unexpectedly long for their molecular complexity. These are targets whose difficulty lies in producing a concise route, rather than finding one at all. The control subset (n = 123) comprises structurally simple molecules that a short, resource-limited AiZynthFinder run nonetheless failed to solve. Because low structural complexity would normally predict an easy synthesis, these unsolved cases are informative, as their failure turns out to be an artifact of the limited search budget rather than a property of the molecules, and under the generous budget used in our benchmark runs most of them are solved. They therefore serve as a control, separating failures caused by an under-resourced search from failures caused by the target lying outside the reachable reaction space.

Together these selections yield 1,098 targets, the benchmark used throughout.

\subsection{The SynthEx multi-agent system}\label{methods:synthex}
SynthEx is an agentic, large-language-model (LLM) retrosynthesis planner
implemented as a subclass of AiZynthFinder's Monte-Carlo tree search
\citep{genheden2020aizynthfinder}, in which the neural template expansion policy
is replaced by an LLM-guided policy.

\paragraph{Strategy generation.}
For each target a \emph{Strategy Generator} first proposes several independent
one-sentence synthetic strategies (default three per target). The prompt directs the
model through a fixed four-point analysis (scaffold motif, key bond-forming
reaction, functional-group conflicts and protection, and stereocenters) and
requests strategies each achievable in one to two reaction steps; the model
returns structured JSON. Chemical constraints (a required starting material, or
free-text human guidance) are optional and are otherwise left to the model's
judgment. Each strategy seeds the subsequent search as a steering query.

\paragraph{Reaction representation and graph edits.}
Disconnections are represented not as reaction SMILES but as an ordered list of
atom-level graph-edit operations (the \texttt{operations} field of an
\texttt{EditBasedRetroReaction}). Each operation is a JSON object keyed by atom-map
number; the operation set comprises ten primitives: \texttt{break\_bond},
\texttt{add\_bond}, \texttt{change\_bond\_order}, \texttt{change\_atom},
\texttt{set\_explicit\_h}, \texttt{add\_group}, \texttt{remove\_group},
\texttt{invert\_stereocenter}, \texttt{clear\_stereocenter} and
\texttt{set\_bond\_stereo}. As a worked example, a retro-Diels--Alder step is
encoded as two \texttt{change\_bond\_order} operations (restoring the diene and
dienophile double bonds) followed by two \texttt{break\_bond} operations (cleaving
the two new sigma bonds). Applying the operation list to the mapped product yields
the mapped precursors deterministically.

\paragraph{LLM configuration.}
The backbone is Gemini (model identifier \texttt{gemini-3.1-pro-preview}) accessed
through the google-genai SDK. Google's published API documentation states a
knowledge cut-off of January 2025 for all Gemini~3 models, including
\texttt{gemini-3.1-pro-preview} \citep{gemini3docs}; we take the cut-off from that
documentation rather than from the model's own report, which is not a reliable
source for it. The cut-off precedes the model's February 2026 release by thirteen
months, leaving a wide window of post-cut-off literature against which the routes
can be compared. The same documentation directs users to the Search Grounding tool
for information after the cut-off; that tool was never enabled here, as set out
below.

Two caveats attach to any cut-off argument. A documented cut-off describes the
pretraining corpus, whereas post-training data is often more recent and is not
generally disclosed. We therefore treat the cut-off as reasonable evidence against
retrieval,
rather than as proof of it. Separately, the target molecules and their scaffolds are
likely present in the pretraining corpus as isolated natural products, together with
discussion of their biosynthesis; our claim is not that the targets are novel to the
model but that no synthetic route to them was available before the cut-off.
Presenting the target structure confers little information on route design, as a
chemist handed the same structure must still plan the disconnections.

No tools were available to the language model at any point. Search grounding was
never enabled and the model had no web access, so it could draw on no information
beyond its training data at inference time. The only lookups in the pipeline are
against the fixed local template library and the building-block stock, both of which
predate the training cut-off.

The strategy planner runs at temperature 0.1 and
the next-step generator at 0.3; no \texttt{top\_p}, \texttt{top\_k} or sampling
seed is set, so generations are not deterministic. Each call is retried up to three
times with a 600\,s timeout. The Phase-1 guided search is bounded by a step limit
of 25. The exact run configuration is
\texttt{synthelite\_config/configs/synthelite.gemini3\_1.code\_ops.yml}.

\subsection{Route criticism and improvement}\label{methods:improve}
Each completed route is serialized as a \emph{RouteJSON} document, a linear sequence
of ReactionJSON entries, so that edits can be applied to individual steps without
regenerating the route. Three agents then operate on it.

The \emph{Critic} agent traverses the route and simulates each reaction in the
forward direction, labeling a step \emph{blocking} when it judges the
transformation chemically infeasible as written, for example because a functional
group elsewhere in the substrate is incompatible with the stated conditions. The
per-route blocking rate reported in Fig.~\ref{fig:route_improve}b is the number of
blocking steps divided by the number of steps in that route.

The \emph{Editor} receives the route together with the \textit{Critic}'s
annotations and resolves the blocking steps by editing the RouteJSON in place. The
permitted operations are reordering steps, inserting or deleting steps, and altering
reaction conditions or functional groups; the agent is instructed to preserve the key
disconnection and the overall strategy. The edited route returns to the \textit{Critic}, and
the loop continues until no blocking steps remain or an iteration cap is reached.

The \emph{Analyst} scores the finished route for overall feasibility on a
five-point scale (\emph{infeasible}, \emph{poor}, \emph{acceptable}, \emph{good},
\emph{excellent}) and identifies its key steps and principal risks
(Fig.~\ref{fig:pipeline}b).

All three agents use the same backbone as the planner. The loop is an internal
consistency procedure, not an external validation: the agent that scores the
improvement shares a backbone with the agents that produce it, so a declining
blocking rate demonstrates convergence against the \textit{Critic}'s criterion and not
experimental feasibility.

\subsection{Search and solve criteria}\label{methods:search}
A building block is deemed purchasable if its full InChIKey is present in the
combined ZINC and eMolecules stock (39,684,411 InChIKeys), matched exactly on the
full InChIKey. A target is scored
as \emph{solved} only when a complete route is returned in which every leaf is
purchasable. This logic is inherited from AiZynthFinder \cite{genheden2020aizynthfinder}.

We report three conditions, all evaluated against the same 1,098-target benchmark and the same ZINC plus eMolecules stock.
\emph{(i) AiZynthFinder exhaustive baseline}: AiZynthFinder run directly on each
intact target with the USPTO expansion, ring-breaker and filter policies, under a
deliberately generous budget (maximum 25 transforms, 1,500 iterations,
1,800\,s wall-clock, returning the first solution); no per-node expansion cap
was set, so the library defaults apply (at most 50 templates per node,
cumulative probability 0.995; exploration constant 1.4). 
\emph{(ii) SynthEx strategic layer only}: a target counts as solved if any
strategy reaches an all-in-stock precursor set using the LLM-guided layer alone,
without template completion. 
\emph{(iii) SynthEx stitched}: any precursor not purchasable after the strategic
layer is completed by a short template search (AiZynthFinder, maximum 6
transforms, 500 iterations, 1,200\,s), and the strategic and template
fragments are stitched into a single route. 

This is a comparison of \emph{reach}, not of compute cost. The LLM planner is
substantially more expensive per target than a template search; the baseline is
therefore run under a generous budget so that its failures reflect the reach of
the USPTO reaction space rather than an exhausted budget. For reference, the
baseline expends a median of approximately $2.9\times10^{4}$ expansion-policy calls
per target (median $\approx 1,500$ MCTS iterations), whereas the short
leaf-completion searches use a median of 163 expansion calls (median 74
iterations); a target with no returned result is counted as an incomplete, hence
unsolved, search.
\subsection{Reaction classification and recognition analysis}\label{methods:classification}
Reactions were classified using three recognition tools, two external and one our own. The first, NameRXN (NextMove Software, \texttt{filbert2} v3.7.0), deems a reaction recognized when it is assigned any named class other than ``Unrecognized.'' The second, Rxn-INSIGHT (\texttt{rxn-insight} v0.1.3), scores a reaction as recognized when it returns a classification other than ``OtherReaction.'' Finally, ReactionClassifier was evaluated in two operating modes: an \emph{Ordered} mode, which applies a hierarchical library of SMIRKS templates and accepts the first matching product, and a \emph{Hybrid} mode, which initially predicts a reaction class using a DRFP fingerprint classifier (2048 bits, radius 2) before applying strict SMIRKS matching restricted to that class's tier-three subset.

ReactionClassifier serves as a purpose-built diagnostic for evaluating distributional distance from the USPTO corpus; by design, it recognizes approximately 57\% of USPTO reactions (57.3\% in Ordered mode). Consequently, any reduced coverage observed on SynthEx chemistry measures the extent to which that chemical space diverges from the underlying USPTO distribution, rather than indicating a technical limitation of the classifier itself. To establish a baseline for this recognition comparison, a USPTO reference set was constructed by sampling 100,000 mapped USPTO reactions (random seed 42), yielding 92,857 parseable, 
single-product reactions after filtering to compare against the 33,145 SynthEx steps.

\subsection{Reaction-space embedding}\label{methods:embedding}
The reaction-space map (Fig.~\ref{fig:space}a) is a principal-component projection
of the output-layer activations of a neural reaction classifier: each SynthEx and
USPTO reaction is passed through ReactionClassifier, its output-layer
representation is taken as a reaction embedding, and the SynthEx and USPTO
embeddings are co-projected by PCA into two dimensions \citep{armstrong2026agentic}. Because that classifier is
itself trained on patent reactions, a degree of separation between the two corpora is
expected on those grounds alone; we therefore present the map as a visualization of
the recognition results rather than as independent evidence of distinctness.

\subsection{Single-step reachability}\label{methods:chimera}
Single-step reachability was assessed with RetroChimera, an ensemble single-step
retrosynthesis model (a SMILES-transformer component and a template-localization
component) using the Pistachio-trained checkpoint. For each SynthEx step, the
model was queried on the largest product fragment and asked for its top 50
predictions. A step counts as \emph{recovered} at rank $k$ if, within the top $k$
predictions, some prediction's set of reactant fragments matches the ground-truth
reactant set. Matching is on canonical SMILES with fragments canonicalized and
sorted; stereochemistry is retained (\texttt{isomericSmiles=True}). Over the full
corpus ($n$~=~33,145) recovery is 13.5\%, 31.4\% and 52.0\% at top-1, top-5
and top-50 respectively; on the ring-forming subset ($n$~=~5,318) it falls to
2.3\%, 10.9\% and 25.8\%.

\subsection{Structural descriptors and ring formation}\label{methods:descriptors}
Structural descriptors were computed per reaction after atom-mapping. A reactant
is treated as a \emph{true} reactant only if it contributes more than three mapped
heavy atoms to the product, which removes reagents, solvents and catalysts; only
single-product reactions are analyzed, and malformed reactions are excluded. A
step is \emph{ring-forming} when the product contains more SSSR rings than the true
reactants combined (RDKit \texttt{CalcNumRings}). Of the 33,145 parseable
SynthEx steps, 5,318 (16.0\%) are ring-forming and 27,827 are not.
Carbon--carbon bond formations are those with NameRXN level-one code 3; among
these, a step is classified as intermolecular when it has two or more true
reactants and intramolecular when it has one. Protecting-group manipulations were
counted as NameRXN level-one codes 5 or 6 (protection and deprotection
combined); on this definition they account for 27.0\% of named SynthEx steps
versus 40.0\% of named RetroChimera disconnections.

\subsection{Atom-mapping and the SynthAtlas release}\label{methods:atlas}
Atom mapping is produced by construction: because each precursor is generated by
applying explicit graph edits to the mapped product, product-derived atoms retain
their parent map numbers and newly introduced atoms are assigned fresh numbers in
sequence. No external atom-mapping model (for example RXNMapper) is used at any point \citep{Schwaller2021mapping}. The released SynthAtlas comprises 1,098 targets with a releasable route, 3,243 strategies
(mean 2.95 per target) and 33,145 reaction steps with valid atom mapping (mean 10.22 steps per route).

\subsection{Expert key-step evaluation}\label{methods:rating}
Individual key steps were rated by ten expert chemists
from academic total synthesis groups in Switzerland and the United States in a blinded web
application. Each item was a single key step; the raters saw the target and
reaction images and a copyable SMILES string, rendered for both sources through a
single RDKit engine with source-neutral filenames and re-canonicalized SMILES, so
that the source (SynthEx or literature) never reached the client; the source
mapping was held only in an offline key. The comparison was conditioned on a shared
strategic frame: from the 70 targets for which both a published literature route
and a SynthEx route to the same target were available, we retained the 47 on
which SynthEx independently arrived at a strategy congruent with the published one
(congruence judged by an LLM), and drew key steps from both routes for these
targets. Key steps were deduplicated by exact canonical reaction SMILES so that no
rater scored two identical steps. Scoring was independent, not a paired forced
choice: each key step was rated on its own on four axes (feasibility, strategic
value, elegance and overall), on a five-point scale with anchors at 1, 3 and
5; the objective axes were presented first and elegance last to limit halo bias,
and ``overall'' was asked as a separate fourth question rather than a composite. In total 1,040 ratings were collected over 148 rated items from ten raters
(per-axis counts, SynthEx/literature: feasibility 672/362, strategic 670/359,
elegance 669/358, overall 672/359).

Agreement between sources on each axis is summarized by Cliff's $\delta$, defined
as $P(\text{SynthEx} > \text{literature}) - P(\text{SynthEx} < \text{literature})$
over per-item mean scores, so that a positive value indicates SynthEx rated higher.
Because each rater scores many items and each item is scored by an uneven number of
raters, ratings are not independent replicates; we therefore treat the rater, not
the rating, as the unit of resampling, and report 95\% confidence intervals from a
cluster bootstrap over the ten raters (2,000 resamples, seed 7), pre-specified as
the primary analysis rather than chosen after inspecting the data. Testing across
the four axes is corrected by a Holm step-down procedure; strategic value is the
only axis whose Holm-corrected interval excludes zero (Section~\ref{res:reach}). On
conventional bands for Cliff's $\delta$, all four effects are negligible to small.
As a robustness check we recompute the comparison rater-by-rater
(Section~\ref{res:reach}), with a leave-one-rater-out sensitivity analysis, and with
Krippendorff's ordinal $\alpha$ for inter-rater reliability\ref{si:tab:reliability}.
The literature key steps were drawn from a set of recent total syntheses published
after the model's training cut-off. From 145 curated routes, 138 yielded a
usable key step; running SynthEx on these targets produced a route to the same
target for 70 of them (the pool above), of which 47 were strategically
congruent. Key steps were extracted by a combination of computational and LLM
analysis: routes were reconstructed and atom-mapped programmatically from the
scraped reactions, and an LLM (Claude Opus 4.8, structured output) then produced a
strategy description and identified the key-step reactions; all routes were
human-confirmed.

\subsection{Statistics, software and reproducibility}\label{methods:repro}
Benchmark construction and the labeling-set build used a fixed random seed
(42); the rating-study
bootstrap uses seed 7 with 2,000 resamples. Analyses used RDKit, AiZynthFinder
\citep{genheden2020aizynthfinder}, NameRXN (\texttt{filbert2} 3.7.0),
Rxn-INSIGHT (\texttt{gen-rxn-insight} 0.1.3) and ReactionClassifier
\citep{namerxn, rxn-insight, armstrong2026agentic}.

The language-model planner is not deterministic: no sampling seed is set, and
temperatures of 0.1 and 0.3 are used for strategy generation and next-step generation
respectively (Methods~\ref{methods:synthex}). Re-running SynthEx on a target will
therefore not reproduce a route exactly. All reported figures come from a single run
over the benchmark.

\textbf{Acknowledgements.}
The authors thank collaborators in EPFL Laboratory of Artificial Chemical Intelligence (LIAC) for helpful discussions.

\textbf{Funding.} This work was supported by the Swiss National Science Foundation through the National Centre of Competence in Research (NCCR) Catalysis (225147) and through the grant (214915). TAN acknowledges support from Intel and Merck KGaA via the AWASES programme. MD acknowledges financial support from the Research Foundation -- Flanders (FWO Vlaanderen) through postdoctoral fellowship grant 1266226N and travel grant V414426N. XVN acknowledges the support from the AiChemist project via MSCA Doctoral Network. GB acknowledges the support from the LowDataML project via MSCA Doctoral Network. This research was conducted with support from Google.org and the Google Cloud Research Credits program for the Gemini Academic Program. PS is part of the Reaxys R\&D collaboration network. NTJ, JF and HL acknowledge support from the US National Science Foundation (US NSF) grant CHE-2449261.

\textbf{Competing interests.} The authors declare no competing interests.

\textbf{Data availability.}

SynthAtlas, comprising 1,098 natural-product targets, 3,243 strategic routes and
33,145 fully specified and atom-mapped reaction steps, is released at
\url{https://synthatlas.epfl.ch} and archived at
Zenodo (DOI to be assigned). The benchmark target list, the AiZynthFinder
baseline results, the reaction-classification outputs underlying
Fig.~\ref{fig:space} and Table~\ref{tab:rings}, and the anonymized expert
key-step ratings underlying Fig.~\ref{fig:results}c are included in the same
database. The building-block stock is derived from ZINC and eMolecules and is
subject to those providers' terms. NPAtlas (release 2024\_09) is publicly
available. Pistachio and NameRXN are commercial and cannot be redistributed.

\textbf{Code availability.} 
The SynthEx framework is released under an open-source license
  at the project repository, \url{https://github.com/schwallergroup/SynthEx.git}.
Deterministic ReactionClassifier is released under an open-source license
  at the project repository, \url{https://github.com/schwallergroup/ReactionClassifier.git}.

\noindent

\bigskip

\bibliography{sn-bibliography}
\appendix

\section{Assembling the blind chemist evaluation set}
We asked expert chemists to score individual synthetic key steps drawn from two
sources: the key steps of recently published academic total syntheses, and the
key steps of the routes proposed by our tool for the same targets. Each chemist
saw one key step at a time and scored it on four scales. The labelers saw the target structure and the reaction diagram. Steps from the two sources
were mixed, blinded and shown in a scrambled order, so raters could not tell which
had come from the literature and which from our tool.

\section{Extracting the key steps}
The two sources are matched by target, but the key step of each route was
identified differently by design. For the literature we identified it
independently, whereas for our tool we used the step the tool itself nominated;
the chemists then scored both on the same scales, so the comparison is between the
steps each source treats as pivotal rather than one assessor's view of both.

For the published syntheses we first reconstructed each route from the scraped
reaction records: reactions were grouped by DOI, atom-mapped, and joined into a
synthesis tree by matching the products of one reaction to the reactants of the
next, allowing for small stereochemical mismatches during matching. We numbered
the reactions of each route in synthetic order, from starting materials to final
target, and gave the numbered route to a large language model (Claude Opus~4.8),
which returned a short description of the strategy and the number of the single
reaction that defines it (occasionally two, for a route that genuinely hinges on a
convergent union followed by a cascade cyclization). That nominated reaction is
the literature key step. Of 145 targets, 138 yielded a key step; the remainder
failed to reconstruct cleanly and were dropped.

\section{Removing routine chemistry}
A central claim of the tool is that it proposes non-obvious disconnections, so we
removed steps that are routine and carry little discriminating signal---protecting-group
manipulations, standard functional-group interconversions, amide and ester
formation, and the like. To do this consistently we classified every candidate
step with a published reaction classifier, which assigns a step to a named class
only when a reaction template reproduces its product exactly. Sixty of the 358
steps were classified in this way; the remaining 83\% were left untouched.

We did not, however, discard every named reaction, because some of the most
strategic steps carry familiar names. We therefore kept classified steps whose
class was a skeleton-forming or olefination reaction, alkynylations, sp$^2$--sp$^2$
couplings, aldol and Horner--Wadsworth--Emmons and Wittig olefinations,
organolithium additions, olefin metathesis and Claisen rearrangements and
removed the rest, which were protecting-group cleavages, oxidations and
reductions, ether, amide, ester, silyl, thioether and $N$-alkyl bond formations,
arene halogenations and Buchwald--Hartwig aminations. This removed 8 literature steps, leaving 156 steps. In addition, we deduplicated reactions across the corpus, removing another 8 steps, for a final count of 148 steps.

\section{Scoring and blinding}
Chemists scored each step on four independent 1--5 scales: feasibility (would the
reaction work as drawn, considering reactivity and selectivity); strategic value
(how much the step builds the target, through the bonds and stereocenters it
forms); elegance (a holistic judgment of the synthetic design); and an overall
``would I run this'' score. A prompt on every item asked raters, when judging
feasibility, to consider whether conditions exist that could achieve the required
chemo-, regio- and stereoselectivity. Each item also allowed a free-text comment
and a flag to mark a step for later exclusion.

Steps were presented without any indication of their source. The mapping from item
to source was held in a separate key that the interface never accessed. Literature
and tool steps were paired and then scrambled within and across pairs so that
neither position nor ordering revealed the source, using a fixed random seed so
that the whole set can be reproduced exactly.

\subsection{Efficiency of the expansion policies}
\label{si:efficiency}

\begin{figure}[t]
  \centering
  \includegraphics[width=\textwidth]{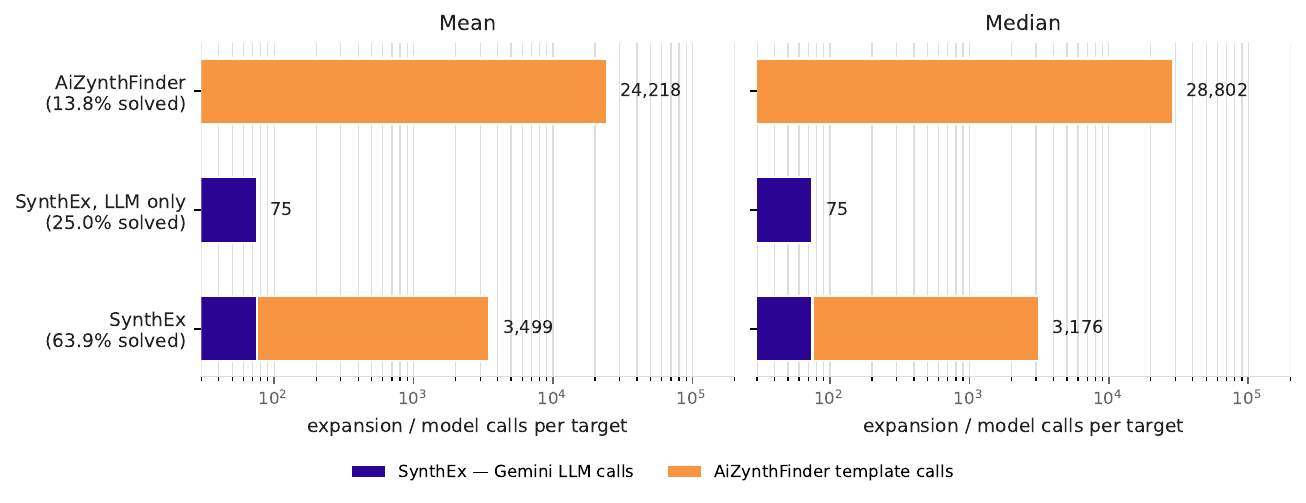}
  \caption{\textbf{Expansion/model calls per target}, shown as the mean (left)
  and median (right) over the full corpus ($n=1098$); solve rate annotated beside
  each method. AiZynthFinder run on the intact target is compared with SynthEx's
  LLM policy alone and the full pipeline, in which AiZynthFinder completes
  SynthEx's unsolved leaves. LLM and template calls are different units of work;
  the bars count invocations, not compute.}
  \label{si:fig:eff}
\end{figure}

The two planners arrive at comparable coverage through very different amounts of
search. Run on the intact target, AiZynthFinder expands tens of thousands of
nodes per molecule---a mean of $24{,}218$ calls to its expansion policy, median
$28{,}802$ (Fig.~\ref{si:fig:eff}). Most of that effort goes into failures: a
target it cannot solve simply runs out the $1500$-iteration budget, whereas the
ones it does solve usually finish in a few hundred calls. SynthEx's LLM policy
works under a hard budget of $75$ calls per target (three strategies, $25$
expansions each) and reaches a slightly higher raw solve rate ($25.0\%$ vs.\
$13.8\%$), roughly two to three orders of magnitude fewer policy invocations for
the same job.

Completing SynthEx's unsolved leaves with AiZynthFinder is where the template
planner is at its best: the leaves are small, near-terminal fragments, and when
one is solvable AiZynthFinder finds a route almost immediately (median $6$
expansion calls). What a target actually costs at this stage is set by the leaves
AiZynthFinder \emph{fails} on---each of those runs out its budget at around
$1{,}650$ calls---so, summed over a target's handful of leaves, the stitch adds a
mean of ${\sim}3{,}400$ template calls (median ${\sim}3{,}100$). Even so, the full
pipeline sits an order of magnitude below the cost of running AiZynthFinder on the
intact molecule, and solves far more of them. One caveat when reading
Fig.~\ref{si:fig:eff}: a Gemini expansion and an AiZynthFinder template call are
not the same unit of work---one is a single large-model inference, the other a
small template-network pass---so the bars count how often each policy is queried,
not the underlying compute.

\subsection{Route length on jointly solved targets}
\label{si:routelen}

\begin{figure}[t]
  \centering
  \includegraphics[width=\linewidth]{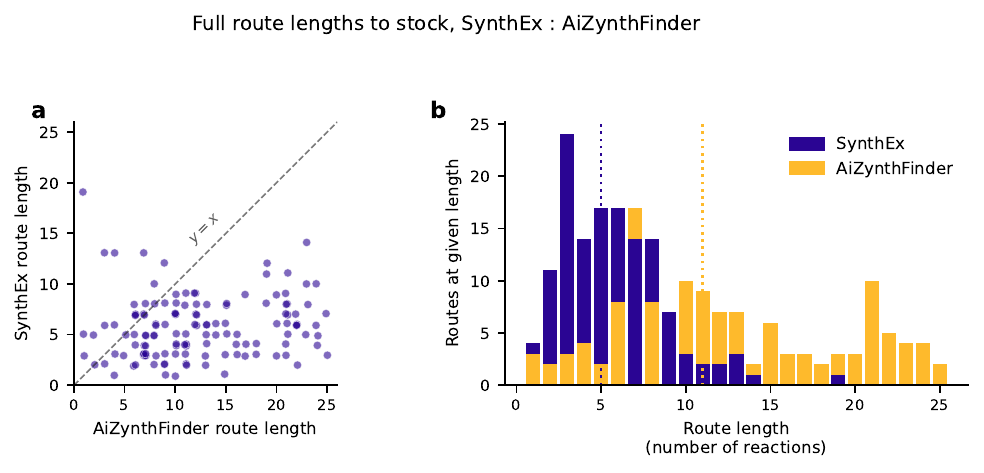}
  \caption{\textbf{SynthEx plans shorter routes.} Full route length (to
  purchasable building blocks) for SynthEx versus AiZynthFinder on the $n=134$
  targets both methods solve. Left: per-target comparison; most points fall
  below $y=x$ ($105/134$). Right: route-length distributions (dotted lines:
  medians, SynthEx $5$ vs AiZynthFinder $11$).}
  \label{si:fig:routelen}
\end{figure}

Where both methods reach a target, SynthEx's routes are shorter. Restricting to
the $n=134$ targets for which both SynthEx and AiZynthFinder return a complete
route to purchasable building blocks, SynthEx's median full route length (its own
strategic disconnections plus any AiZynthFinder-completed leaves) is $5$ steps
(mean $5.7$) against AiZynthFinder's median $11$ steps (mean $12.2$)
(Fig.~\ref{si:fig:routelen}). SynthEx reaches the target in fewer steps on $105$ of
the $134$ targets ($78\%$).

\section{Robustness of the expert key-step comparison}
\label{app:ratings_robustness}

The main text (Fig.~4c) reports the four-axis comparison between SynthEx and
literature key steps as a cluster bootstrap over ten raters, the appropriate
resampling unit since each rater scores many items and each item is scored by an
uneven number of raters. This section reports that comparison decomposed by
participating group, its dependence on any single rater, and the reliability of
the rating instrument itself.

\begin{figure}[htbp]
\centering
\includegraphics[width=\textwidth]{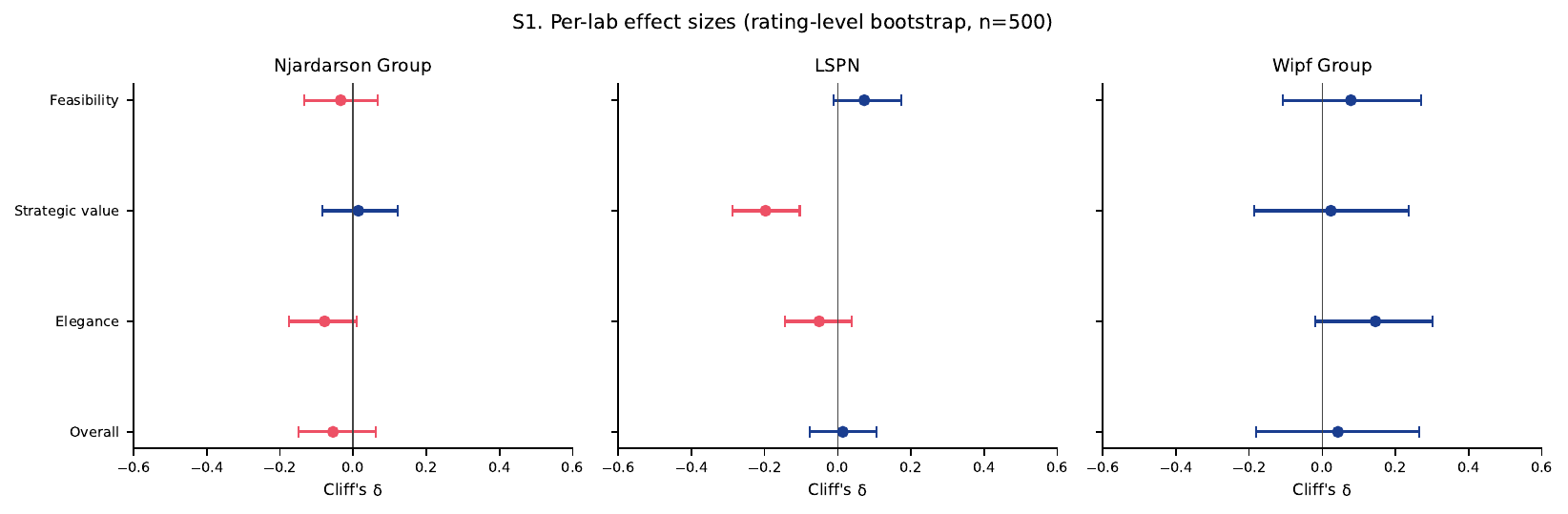}
\caption{\textbf{Per-group effect sizes.} Cliff's $\delta$ (SynthEx vs.\
literature) by rating axis, computed separately within each participating group
(rating-level bootstrap, 500 resamples). LSPN is the only group whose interval
excludes zero on any axis (strategic value); the Njardarson Group and the Wipf
Group show no detectable difference on any of the four axes.}
\label{si:fig:forest}
\end{figure}

Figure~\ref{si:fig:forest} decomposes the comparison by participating group. The
pattern underlying the pooled strategic-value effect is uneven across groups
rather than uniform: LSPN's own ratings alone give $\delta = -0.20$ (95\% CI
[$-0.29$, $-0.10$]), the only per-group interval on any axis that excludes zero,
while the Njardarson Group ($\delta = +0.01$, [$-0.08$, $+0.12$]) and the Wipf
Group ($\delta = +0.02$, [$-0.19$, $+0.24$]) show no detectable difference on the
same axis. No group shows a detectable difference on feasibility, elegance or
overall quality.

\begin{figure}[htbp]
\centering
\includegraphics[width=0.6\textwidth]{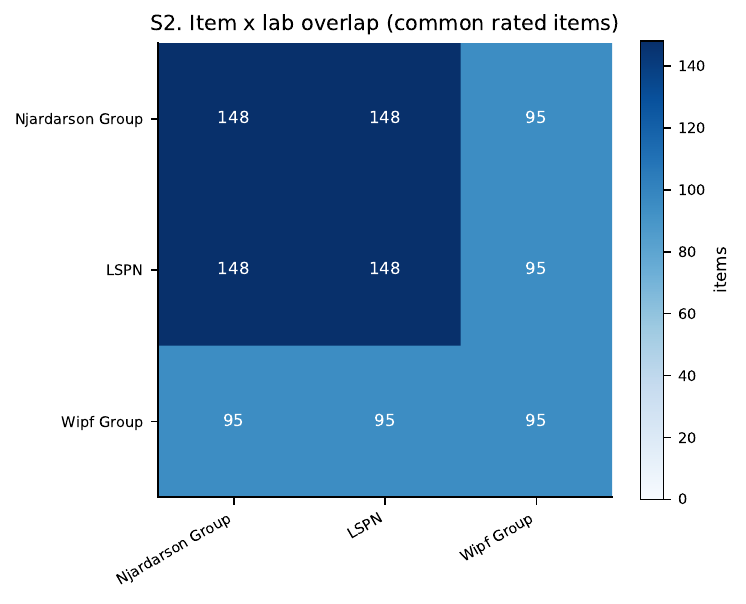}
\caption{\textbf{Item-by-group overlap.} Number of key steps rated in common by
each pair of groups. The Njardarson Group and LSPN rated the full 148-item set
each; the Wipf Group's 95 items are a strict subset rated identically by both
other groups, so the per-group effect sizes in Fig.~\ref{si:fig:forest} are not
confounded by item difficulty.}
\label{si:fig:overlap}
\end{figure}

This unevenness is not an artifact of the groups rating different material.
Figure~\ref{si:fig:overlap} shows the number of key steps each pair of groups
rated in common: the Njardarson Group and LSPN rated the full item set (148 of
148 items each, entirely overlapping), and the Wipf Group's 95 items are a strict
subset rated identically by both other groups. The three groups' effect sizes are
therefore directly comparable and are not confounded by item difficulty.

\begin{figure}[htbp]
\centering
\includegraphics[width=\textwidth]{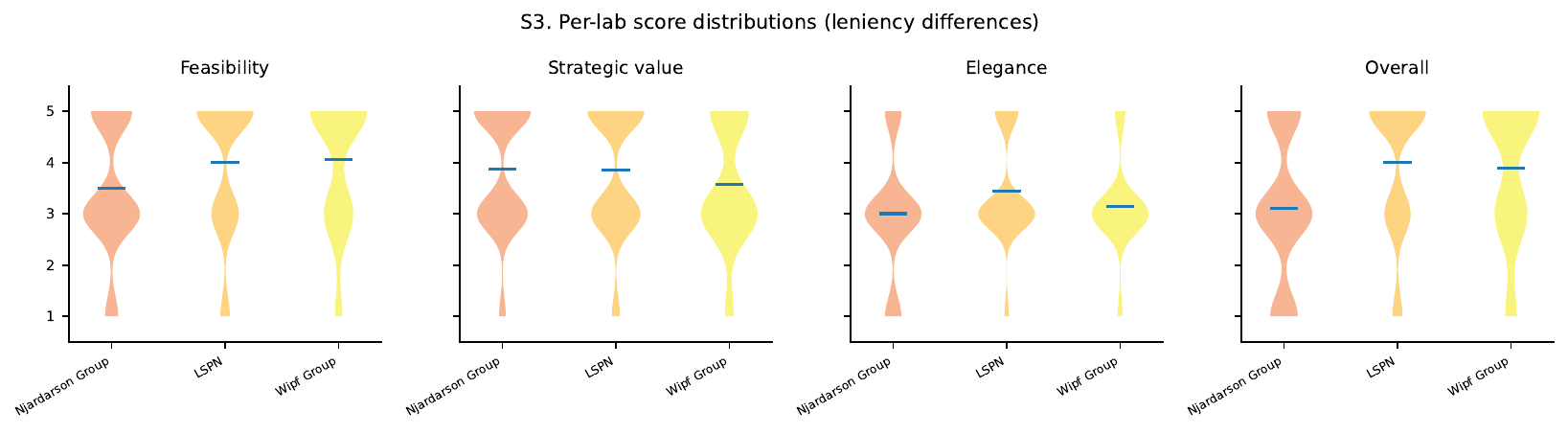}
\caption{\textbf{Per-group score distributions.} Raw score distributions (both
sources pooled) by rating axis and participating group, showing that groups
differ in overall scale use. This is why every effect size in this section and
the main text is computed within rater or within group, never from raw means
compared across raters directly.}
\label{si:fig:leniency}
\end{figure}

Figure~\ref{si:fig:leniency} shows each group's raw score distribution by axis.
Groups differ noticeably in overall scale use (some are more lenient across both
sources than others), which is why every effect size in this section and the
main text is computed within rater or within group rather than by comparing raw
means across raters directly.

\begin{figure}[htbp]
\centering
\includegraphics[width=\textwidth]{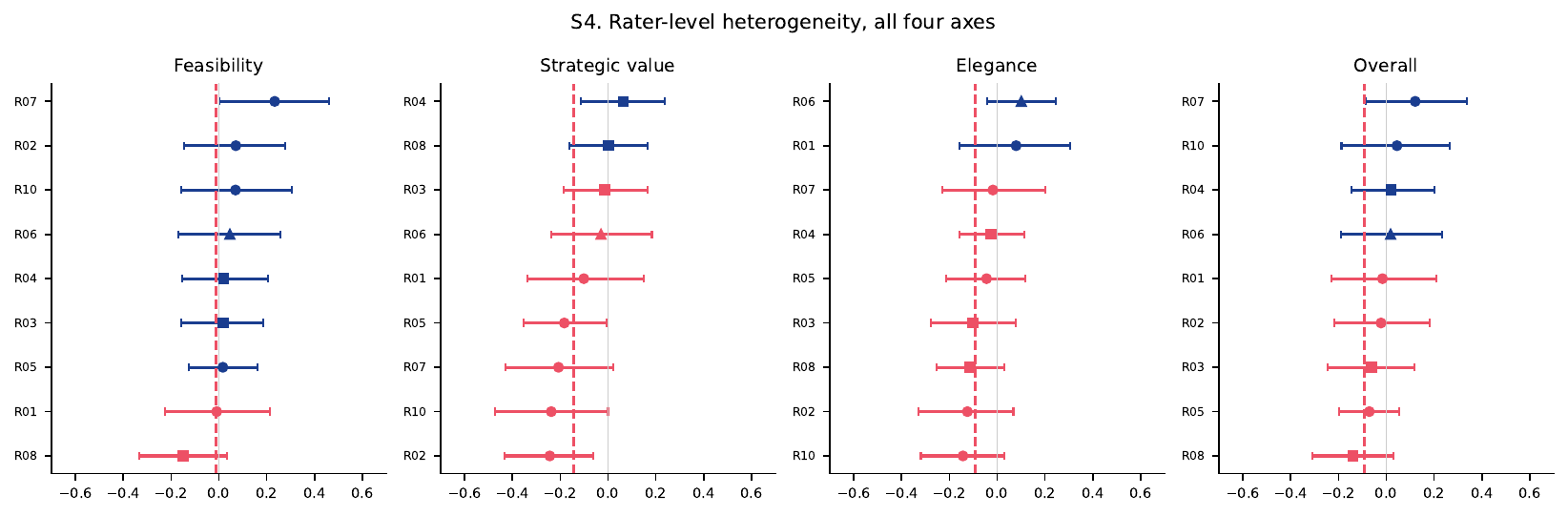}
\caption{\textbf{Rater-level heterogeneity, all four axes.} Individual raters'
own Cliff's $\delta$ (SynthEx vs.\ literature, computed from each rater's own
scores only), extending main text Fig.~4d to feasibility, elegance and overall
quality. Marker shape indicates participating group (squares: Njardarson Group;
circles: LSPN; triangles: Wipf Group); color indicates the direction of that
rater's own estimate (blue: SynthEx-favoring; pink: literature-favoring). The
dashed line is the pooled estimate for that axis. On every axis, at least one
rater's own comparison falls on the opposite side of zero from the pooled
estimate.}
\label{si:fig:caterpillar}
\end{figure}

Figure~\ref{si:fig:caterpillar} extends the rater-level heterogeneity view (main
text Fig.~4d) to all four axes. The spread of individual raters' own estimates
exceeds the pooled effect on every axis (main text), and this figure shows the
pattern is not specific to strategic value: on every axis, at least one rater's
own comparison sits on the opposite side of zero from the pooled estimate.

Table~\ref{si:tab:loo} reports the pooled effect size recomputed with each rater
removed in turn, using the same cluster bootstrap as the primary analysis. The
maximum shift from dropping any single rater is $0.068$ (feasibility), $0.033$
(strategic value), $0.065$ (elegance) and $0.037$ (overall); on strategic value,
the effect remains negative under every possible single-rater removal
($\delta \in [-0.168, -0.109]$). No axis is materially dependent on any one
rater.

Table~\ref{si:tab:reliability} reports inter-rater reliability (Krippendorff's
$\alpha$, ordinal metric, on the 148 items with three or more raters) for each
axis: $0.23$ for feasibility, $0.24$ for strategic value, $0.13$ for elegance and
$0.17$ for overall. Values below $0.4$ are conventionally considered poor
agreement, indicating that individual raters agree with each
other only weakly on the absolute quality of a given step, independent of
source. This is the same phenomenon quantified directly in the main text (rater
heterogeneity exceeding the pooled source effect on every axis) and is the
reason the comparison is reported at the level of a bootstrap over raters rather
than a single pooled estimate treated as if it had no clustering structure.

\begin{table}[htbp]
\centering
\caption{\textbf{Leave-one-rater-out.} Pooled Cliff's $\delta$ (SynthEx vs.\
literature; positive favors SynthEx) recomputed with each rater removed in
turn, using the same cluster-bootstrap point estimate as the primary analysis.
\textbf{None} is the full-panel estimate (main text Fig.~4c). No axis is
materially dependent on any one rater; strategic value remains negative under
every single-rater removal.}
\label{si:tab:loo}
\begin{tabular}{lcccc}
\toprule
Rater dropped & Feasibility & Strategic value & Elegance & Overall \\
\midrule
R01 & $-0.011$ & $-0.137$ & $-0.128$ & $-0.082$ \\
R02 & $-0.011$ & $-0.109$ & $-0.079$ & $-0.069$ \\
R03 & $-0.007$ & $-0.164$ & $-0.089$ & $-0.088$ \\
R04 & $-0.003$ & $-0.168$ & $-0.108$ & $-0.102$ \\
R05 & $-0.012$ & $-0.128$ & $-0.112$ & $-0.083$ \\
R06 & $-0.012$ & $-0.144$ & $-0.123$ & $-0.099$ \\
R07 & $-0.034$ & $-0.136$ & $-0.109$ & $-0.095$ \\
R08 & $+0.056$ & $-0.146$ & $-0.026$ & $-0.053$ \\
R09 & $-0.010$ & $-0.149$ & $-0.098$ & $-0.087$ \\
R10 & $-0.025$ & $-0.125$ & $-0.064$ & $-0.106$ \\
\midrule
\textbf{None} & $\mathbf{-0.011}$ & $\mathbf{-0.142}$ & $\mathbf{-0.091}$ & $\mathbf{-0.091}$ \\
\bottomrule
\end{tabular}
\end{table}

\begin{table}[htbp]
\centering
\caption{\textbf{Reliability.} Ordinal Krippendorff's $\alpha$ per rating axis,
computed over the 148 items with three or more raters. Values below $0.4$ are
conventionally rated poor (Krippendorff, 2004): raters agree only weakly with
each other on the absolute quality of a given step, independent of source,
which is why the main text reports effect sizes with a cluster bootstrap over
raters rather than treating the panel as a single low-variance measurement.}
\label{si:tab:reliability}
\begin{tabular}{lcc}
\toprule
Axis & Krippendorff's $\alpha$ & $n$ items \\
\midrule
Feasibility & $0.233$ & 148 \\
Strategic value & $0.238$ & 148 \\
Elegance & $0.131$ & 148 \\
Overall & $0.174$ & 148 \\
\bottomrule
\end{tabular}
\end{table}

\end{document}